\documentclass[%
reprint,
amsmath,amssymb,aps
]{revtex4-1}

\usepackage{graphicx}
\usepackage{dcolumn}
\usepackage{bm}
\usepackage{color}
\usepackage{epstopdf}
\usepackage{amsmath}

\usepackage[
total={6.5in,8.75in}, top=1.5in, left=0.9in, 
]{geometry}

\usepackage{hyperref}

\begin{document}

\preprint{APS/123-QED}

\title{The new face of multifractality: Multi-branchedness and the phase transitions in time series of mean inter-event times}

\author{Jaros{\l}aw Klamut}
\email{Jaroslaw.Klamut@fuw.edu.pl}
\affiliation{%
Faculty of Physics, University of Warsaw, Pasteur Str. 5, PL-02093 Warsaw, Poland
}%
\author{Tomasz Gubiec}
\affiliation{%
Center for Polymer Studies and Department of Physics, Boston University, Boston, MA 02215 USA
}%
\affiliation{%
Faculty of Physics, University of Warsaw, Pasteur Str. 5, PL-02093 Warsaw, Poland
}%
\author{Ryszard Kutner}
\affiliation{%
Faculty of Physics, University of Warsaw, Pasteur Str. 5, PL-02093 Warsaw, Poland
}%
\author{Zbigniew R. Struzik}
\affiliation{%
University of Tokyo, Bunkyo-ku, Tokyo 113-8655, Japan
}%
\affiliation{%
Advanced Center for Computing and Communication, RIKEN, 2-1 Hirosawa, Wako 351-0198, Saitama, Japan
}%


\begin{abstract}
Empirical time series of inter-event or waiting times are investigated using a modified Multifractal Detrended Fluctuation Analysis operating on fluctuations of mean detrended dynamics. The core of the extended multifractal analysis is the non-monotonic behavior of the generalized Hurst exponent $h(q)$ -- the fundamental exponent in the study of multifractals. The consequence of this behavior is the non-monotonic behavior of the coarse H\"older exponent $\alpha (q)$ leading to multi-branchedness of the spectrum of dimensions. The Legendre-Fenchel transform is used instead of the routinely used canonical Legendre (single-branched) contact transform. Thermodynamic consequences of the multi-branched multifractality are revealed. These are directly expressed in the language of phase transitions between thermally stable, metastable, and unstable phases. These phase transitions are of the first and second orders according to Mandelbrot's modified Ehrenfest classification. The discovery of multi-branchedness is tantamount in significance to extending multifractal analysis. 

\begin{description}

\item[PACS numbers]
89.65 Gh, 05.40.-a, 89.75.Da
\end{description}
\end{abstract}

\maketitle



\section{Introduction}\label{section:Introduction}

The concept of extended scale invariance, referred to as multifractality, has become a routinely applied but still intensively developing methodology for the study of both complex many-body systems~\cite{OKD,KD,ODFJK,DO,BBM1,CHC} and nonlinear (e.g., chaotic with a low degree of freedom) dynamical systems~\cite{BeSch}. 
It is a rapidly evolving and inspiring approach to nonlinear science in many different fields, stretching far beyond traditional physics~\cite{JXZS}.

\subsection{Remarks on multi-branched multifractality}\label{section:l-smulti}

Only two sources of true multifractality have been identified to date~\cite{KZKBHS}: (i) the presence of a broad (long/heavy tailed) distribution in the system and/or (ii) the presence of long-term/range dependence (e.g., nonlinear long-term/range correlations), leading to the hierarchical organization of many scales. Moreover, there is a widespread belief that some stochastic or deterministic nonlinear mixture of monofractals should produce multifractals~\cite{BBM, BeSch,SchL}. All of them can create cascades that lie at the heart of multifractality. The (true) multifractality occurs where fluctuations and/or dependences arise in many different spatial and/or temporal scales under different scaling laws, i.e. defined by various scaling exponents, which create a multiscaling phenomenon. Unfortunately, the physical origin of multifractality is, in fact, rarely identified.

The direct inspiration of the present work is drawn from our earlier results presented in Refs.~\cite{PMKK,KKPM}. In these works, we found left-sided multifractality on financial markets as a direct result of non-analytic behavior of the R\'enyi exponent. We encounter left-sided multifractality when the spectrum of dimensions (or singularities) extends only to the left of its extreme value. This implies that the spectrum of dimensions, which in the standard case is roughly the shape of a horseshoe with arms turned down, has only the left arm. Previously, we found that a specific complex, broad distribution of inter-event times is responsible for the existence of left-sided multifractality. It is a result of the Mixture of Distribution Hypothesis in finance~\cite{KKPM} (and refs. therein). In the present work, we suggest, however, that nonlinear long-term autocorrelations bear the primary responsibility for the multifractality observed.

Two essential elements of the multifractality considered in the present work are: (i) the dominant left-hand asymmetry of the spectrum of dimensions, i.e., the so-called left-sided multifractality, favoring large fluctuations, and (ii) an entirely novel concept of multi-branched multifractality. We deal with multi-branched multifractality when the spectrum of dimensions is multi-branched where the concept of 'branch' is to be further defined below. We can now clarify that in this case, the term `left-sided' refers to the dominant branch (i.e., the one on which the contact point lies). A detail definition of the dominant, main branch is given in Sec.~\ref{section:fsphasetr}, where we illustrate this concept in Fig.~\ref{figure:cqvsq}(b). Item (i) is discussed in Sec.~\ref{section:l-smulti}  below, while item (ii) is discussed in Sec.~\ref{section:multinit}. Therefore, in the present work, we are considering multi-branched multifractality with dominant left-sided character.

Attention was first drawn to the existence of left-sided multifractality by Mandelbrot and coauthors~\cite{ManEvHa,ManEv}. They discovered the existence of a spectrum of dimensions with only one, left-side branch (or arm). This left-sided multifractality was generated by the binomial cascade, which produces singularity in the R\'enyi exponent or stretched exponential decay of the finest coarse-grained probability. 

Blumenfeld and Aharony~\cite{BlAh} discovered an exciting breakdown of multifractality in diffusion-limited aggregation (DLA). They found strongly asymmetric left-sided spectra of singularities depending on the size of the growing aggregate in DLA. Their spectrum shows an apparent tilt to the left as a signature of the phase transition to non-multifractality. 

Earlier, multifractals with the right part of the spectrum of singularities not well defined (effect caused by a phase transition), were mimicked by a random version of the paradigmatic two-scale Cantor set and also in the domain of DLA~\cite{BCJ,LS,JPV,HHD} (and refs therein).

\subsection{Remarks on identification of multifractality}\label{section:multinit}

In recent years much effort has been devoted to reliable identification of the true multifractality in real data. These data are coming from various fields such as geophysics~\cite{SchL}, seismology~\cite{BE} and hierarchical cascades of stresses in earthquake patterns~\cite{MVR,SOu}, from atmospheric science and climatology (e.g., turbulent phenomena~\cite{SLDS,XJJHF}), financial markets~\cite{BBM,HKY}, neuroscience~\cite{KJM} (e.g., neuron spiking~\cite{DF}), from cardiac science or cardiophysics~\cite{CP} (e.g., physiology of the human heart~\cite{FJ} and refs. therein), and from further works investigating complexity in heart rate~\cite{PIB,ZRS} and physiology~\cite{NHE}. However, the identification of true multifractality is still a challenge.

Specifically, the verification of multifractality in empirical data requires caution due to strong non-stationarities, such as crashes~\cite{CzG}, and also because of the presence of spurious~\cite{LBKSB} and/or corrupted multifractality~\cite{SchK}. There are also other difficulties with the identification of the true multifractality, primarily when nonlinear properties of the time series are studied. A spurious multifractality can also arise as a result of slow crossover phenomena on finite timescales~\cite{BPM}. Furthermore, pollution of a multifractal signal with noise (white or colored) as well as the presence of short memory or periodicity can significantly alter the properties of the multifractal signal.

Further, the limited amount of empirical data available and the resulting limited range of physical multifractality is a serious technical challenge. 
The limited range and amount of empirical data can be the source of finite size effects~\cite{GPa}. Fortunately, because multifractality is extremely sensitive to these contaminating effects, they can be appropriately identified and eliminated or at least minimized. In our case we deal with the situation where the role of the finite-size effect is small compared to other factors.

Last but not least, identifying true multifractality is difficult because we are not sure that all the sources of multifractality have been discovered to date~\cite{KZKBHS}.

Clearly, the situation is complicated. Nevertheless, in our work we demonstrate, by studying the time series of inter-event times, that the extraction of true multifractality is possible. The above-mentioned difficulties in detecting true multifractality are a call to action for developing powerful multifractality detection methods such as the method presented here. We also aid under-researched inter-event times analysis methodologies gaining importance as an essential element of the modern continuous-time random walk formalism~\cite{MoMa,TM,MMSZ}. 

\subsection{Work purpose overview}

The use of the empirical series of inter-event times by us is of a generic nature. It is a characteristic example of the time series for which generalized Hurst and coarse H\"older exponents, as well as R\'enyi dimensions, exhibit non-monotonic behavior vs. order of scale. The study of the consequences of this non-monotonicity is one of the exciting subjects of this work.

Financial markets fluctuate, sometimes strongly by increasing the risk level in order to maximize profit. This finds its reflection in the inter-event times' patterns acting as a direct reflection of the systems' activities -- their various properties were studied in the last decade~\cite{OKD2,OKD,OKR,PMKK,KKPM,DGKJS,ODGGJ,JHKK,BG}. Among them, the key observation is that quite often the dependence between waiting times dominates that between spatial increments~\cite{KGK} defining the process, which cannot be considered as a renewal process~\cite{EFel}. Without examining the role of inter-event times, we are not able to describe the dynamics of financial markets -- this examination is still at an early stage of development. This situation is the motivation and inspiration for our work, emphasizing the crucial role as mentioned above of inter-event times. It is essential, however, to realize that the generic goal of the work is to significantly expand the multifractal methodology to be capable of inferring true multifractality. This capacity of our methodology achieved here motivates us to label it as a 'new face' of multifractality and formalisms of its investigation. Indeed, our contribution is primarily of such neoteric methodological character.

To this aim, we study empirical fluctuations of inter-event times and their mutual dependencies by relying on their absolute central moments and autocorrelations of fluctuations' absolute values. In the case of financial markets, the fluctuations are generally speaking a consequence of the double-auction mechanism~\cite{ES1,ES2,PS}, where different types of orders compete with each other. This approach allows ordering the fluctuations according to the degree of their corresponding moments (cf. the Lyapunov inequality in Ref.~\cite{Lyapunov}). It is essential in a multiscaling analysis in many branches of science. 

The canonical multifractal detrended fluctuation analysis (MF-DFA) is a reference approach. However, our approach differs from it in several essential points. For example, we correctly take into account the normalized partition function (which is guaranteed to be non-negative in our approach, as it should be). This partition function is built based on the normalized fluctuation function.

We reveal multi-branched multifractality, where the first and second-order phase transitions exist together with both thermal stable and unstable phases. The stable phase is defined by the positive value of the specific heat, and the unstable phase by its negative value. This is discussed in Sec.~\ref{section:fsphasetr} and illustrated in Fig.~\ref{figure:cqvsq}. It remains a challenge to find microscale physical mechanisms (or at least surrogates) underlying multi-branched multifractality discovered. We expect this discovery to play a significant role in the future analysis of the real-time series of different origins e.g., geophysical, medical, and financial time series.

More specifically, the non-monotonic behavior of the generalized Hurst exponent which we found results in turning points in the plot of the coarse H\"older exponent. It is directly responsible for the multi-branched spectrum of dimensions and for the first and second-order phase transitions together with thermally stable and unstable multifractal phases. For the analysis of the multi-branched spectrum of singularities on the financial market, the application of the Legendre-Fenchel contact transform is necessary (in a way complementary to that used in Ref.~\cite{PMKK,KKPM}). This transform is a generalization of the canonical Legendre contact transform routinely used to extract usual single-branched multifractality from empirical data. Generally speaking, the Legendre-Fenchel transform allows many solutions, rather than the single solution permitted by the Legendre transform. Interestingly in this context, a slight non-monotonic behavior of the generalized Hurst exponent has  recently been observed  on the Bitcoin (BTC) market for BTC prices~\cite{DGMOW}.

It needs to be highlighted and motivated that we decided to develop an analysis method belonging to the DFA group (see review~\cite{CHC} and refs. therein) and not to the coarse-graining group for reasons presented in Refs.~\cite{OKD,DGZM}. These works compare the effectiveness of MF-DFA both with Wavelet Transform Modulus Maxima (WTMM) and with Detrended Moving Average (DMA), two canonical representatives of coarse-graining methods. Ref.~\cite{OKD} concludes, that in the majority of situations in which one does not know a priori the fractal properties of a process, choosing MF-DFA should be recommended instead of the WTMM. On the other hand, Ref.~\cite{DGZM} corroborates that the DMA method gives an over-estimation of the Hurst exponent in comparison with the DFA technique. However, our method involves pre-processing based on averaging which is equivalent to the coarse graining approach.

It is worth paying attention to one more matter. For data containing intrinsic trends, DFA methods are usually required. The time series of intraday inter-event times used in this work contain trends of this type. Such intrinsic trends are, among others, caused by the so-called \emph{lunch effect}~\cite{GubWil}. From these increments, the profile, or walk, is built to which DFA techniques can be directly applied, as was done e.g., in Ref.~\cite {OKD}. More specifically, the extended MF-DFA method is developed in this work, exploring the multi-branched multifractal character of long-term autocorrelated intra-day time intervals. The multi-branched multifractal is a novel concept expressed in terms of both continuous and discontinuous phase transitions and summarized and termed as 'the new face of multifractality'.

The organization of the work is as follows. In the present~Sec.~\ref{section:Introduction}, above we give the motivation of our work and its goal, indicating a possibility of extension of our approach to research areas far beyond the specific example of financial time series used here. In Sec.~\ref{section:NMF-DFA}, the extension of the canonical MF-DFA is developed and applied to the description of the to date insufficiently exploited empirical time series of inter-event times. In Sec.~\ref{section:fsphasetr}, we reveal the existence of the first and second-order phase transitions in this type of multifractality and examine the main thermodynamical consequences. Finally, in Sec.~\ref{section:disconc}, we discuss critical results of the work, indicate their importance, and summarize the whole work. The main body of the work is supplemented with Appendices A through H, providing additional discussion of the selected points deserving attention.

\section{Normalized Multifractal Detrended Fluctuation Analysis}\label{section:NMF-DFA}

The main subject of this work is the analysis of the true multifractality generated by the non-monotonic behavior of the generalized Hurst exponent. This non-monotonicity manifests itself in the multi-branch spectrum of dimensions. Specifically, the multi-branch spectrum of dimensions we identify belongs to the class of multifractality with a dominant left-sided branch. 

Our approach combines statistical-physical analysis, based on the generalized statistical-mechanical partition function, with that based on the multiscale fluctuation function. It takes us from the absolute moments of arbitrary orders through the partition function to multifractality. 

The central role in the analysis we develop is taken by the generalized Legendre transform -- the Legendre-Fenchel transform, which is also referred to as the generalized contact transform. The weak non-monotonic $q$-dependence of the generalized Hurst exponent was already observed in both real and spurious multifractality contexts in Ref.~\cite{RRDG} (and refs. therein) -- however, its consequences have not been studied to date.

Further, we develop a normalized extension to the standard multifractal detrended fluctuation analysis (NMF-DFA) ready for the study of both stationary and non-stationary detrended time series. In particular, we allow that after detrending, time series may still contain some higher-order non-stationarities which are then properly addressed. This is possible due to the consistent definition of the probability of fluctuations introduced in Sec.~\ref{section:GPF} by Eq. (\ref{rown:pjnus}). 
This probability of fluctuations, to be reffered to as escort probability, is more adequate than the non-normalized and sometimes even negative probability given by Eq. (12) in Ref.~\cite{KZKBHS}. 

We are dealing only with the analysis of detrended absolute values, i.e., these bereft of dichotomous noise. The motivation is that this type of nonlinear quantities can be long-term autocorrelated as opposed to the (usual) bilinear autocorrelations.
In our case, the autocorrelations which are studied point to the existence of a distinct antipersistent structure of fluctuations behind them. We hypothesize, that this structure reflects the fact that after a period of high market activity, there is a period of significantly lesser activity and so on in an alternating fashion, leading to the effect of volatility clustering.

\subsection{Intra-day fluctuations of inter-event times: pre-processing of our formalism}\label{section:ifit}

The entire time series of inter-event times is naturally divided into $N_d$ trading days or sessions of equal duration $T$. Durations of weekends are not included in the time series. The end of the trading week is on Friday and the trading week starts again the following Monday. Similarly, the duration times of night breaks are not included. Further, each trading session is divided into $s$ non-overlapping daily time windows of equal duration $\Delta $. The dimensionless number $s$ defines the daily timescale in our approach. Hence, we have $T=s\cdot \Delta $, where both $N_d$ and $T$ are independent of the scale $s$. 

\begin{figure*}
\begin{center}
\includegraphics[scale=1.10,angle=0,clip]{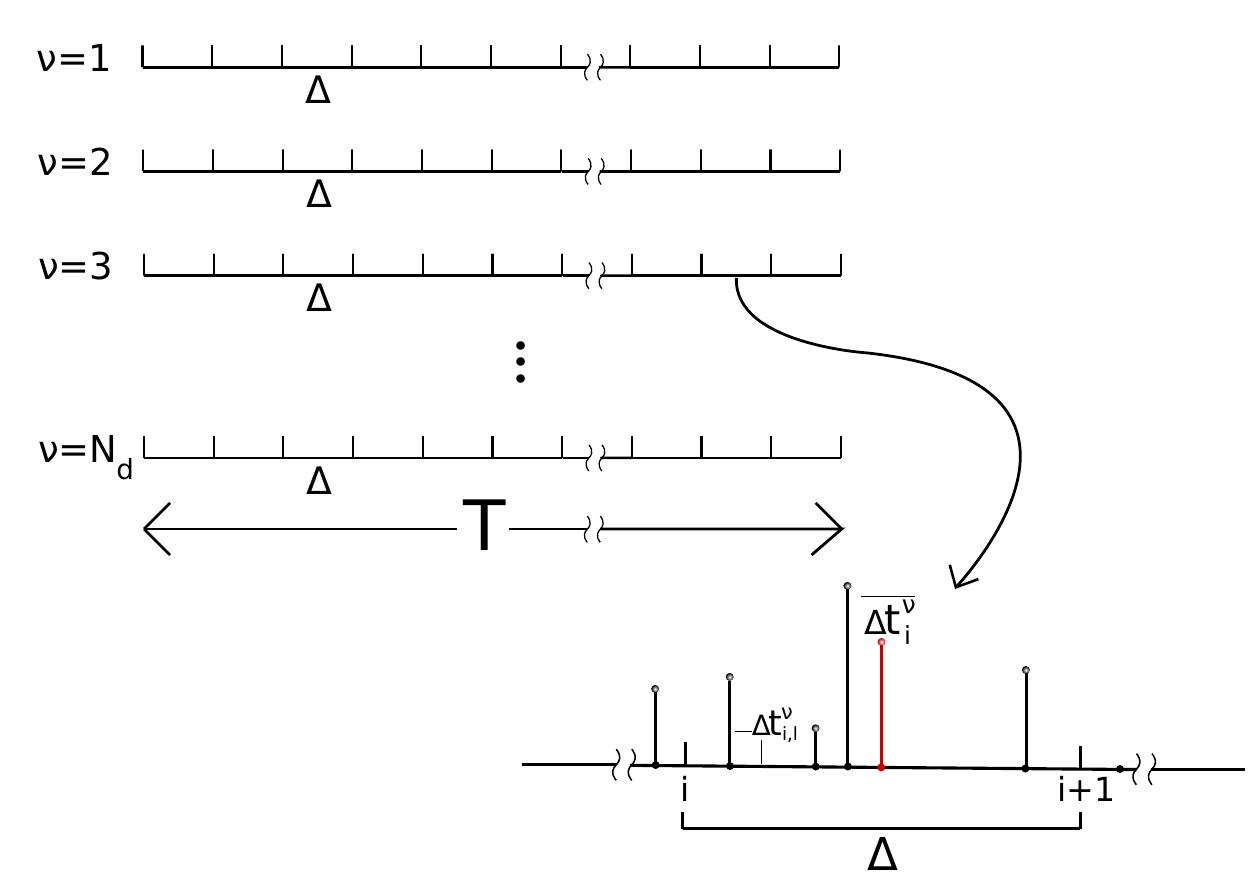}
\caption{Schematic diagram defining inter-event times, time windows, and corresponding means. 
The mean for $i$th time window, $[i,i+1[,\; 1\leq i\leq s$, (of $i$-independent width $\Delta $) is given by the corresponding time average $\overline{\Delta t_{i}^{\nu }}$. We define the inter-event time  $\Delta t_{i,l}^{\nu }$ as belonging to $i$th time window $[i,i+1[$, when at least the left border of inter-event time $\Delta t_{i,l}^{\nu }$ belongs to it (obviously, we have $1\leq l\leq n_i^{\nu }$). Hence, the inter-event time between the final transaction on Friday and the first transaction on the following Monday is assigned to the last time window on Friday. Similarly, we exclude weekend closing time and nights between successive trading days. Focusing on ranges of $s\geq 1$ (keeping $n_i^{\nu }\geq 1$ for any $i$ and $\nu $) we can write $\overline{\Delta t_{i}^{\nu }}\approx \Delta /n_i^{\nu }$. In the particular example of this schematic diagram, $n_i^{\nu }=4$. Thus, our time series consists of local average values shown by red time intervals in the enlarged bottom plot of the time interval $\Delta $.}  
\label{figure:scheme}
\end{center}
\end{figure*}

The rudimentary Fig.~\ref{figure:scheme} shows the construction of the inter-event time series in detail. Our time series consists of local average values -- red time intervals shown in the enlarged bottom drawing of the time interval $\Delta $. In the original work introducing the MF-DFA~\cite{KZKBHS}, the authors used a different notation because the entire time series was there segmented differently.

Our division of the time series essentially distinguishes our approach from the standard MF-DFA used so far (cf. Appendix~\ref{section:AppendA}). The time series we use have double indexation compared to their indexation under the canonical MF-DFA method. This is the basic difference between our method and the standard canonical MF-DFA. We consider its important consequences in Sec.~\ref{section:GPF}. These consequences relate to intra-day properties of time series.

The intra-day (nonlinear) autocorrelation of the absolute additively detrended profile is defined for a single trading day $\nu ,~1\leq \nu \leq N_d$, and within a timescale $s$,
\begin{eqnarray}
&F^2&(j;{\nu },s) \nonumber \\
&=&\frac{1}{s-j}\sum_{i=1}^{s-j} \mid U_{\nu }(i)-y_{\nu }(i)\mid \nonumber \\
&\cdot &\mid U_{\nu }(i+j)-y_{\nu }(i+j)\mid ,\; \nu =1,\ldots ,N_d,
\label{rown:F2snj}
\end{eqnarray}
where dimensionless index $i(=1,2,\ldots ,s)$ enumerates the current time window of length $\Delta $. The dimensionless index $j(=0, \ldots, s-1)$ defines time-step distance or the number of time windows of width $\Delta $ between absolute deviations (detrended fluctuations) $\mid U_{\nu }-y_{\nu }\mid $ present at day $\nu $ at time windows $i$ and $i+j$; function $y_{\nu }$ is the detrending polynomial, while daily profile $U_{\nu }$ is defined by Eq. (\ref{rown:Yy}) below. 

As is evident from Eq. (\ref{rown:F2snj}), we take into account only intra-day absolute autocorrelations, separately for each day. This implies that detrending polynomials are fitted separately for each trading day of a fixed duration time for any timescale. This is an essential difference from the canonical MF-DFA, where the duration times of the segments change with the timescale. Our way of detrending avoids possible artificial fluctuations at the ends of any segment inside trading days. The average values obtained within the statistical ensemble of the trading days are subject to a relatively small statistical error. This approach allows us to analyze the subtle effect of the non-monotonic characteristics of multifractality.

\begin{figure*}
\begin{center}
\includegraphics[scale=0.65,angle=0,clip]{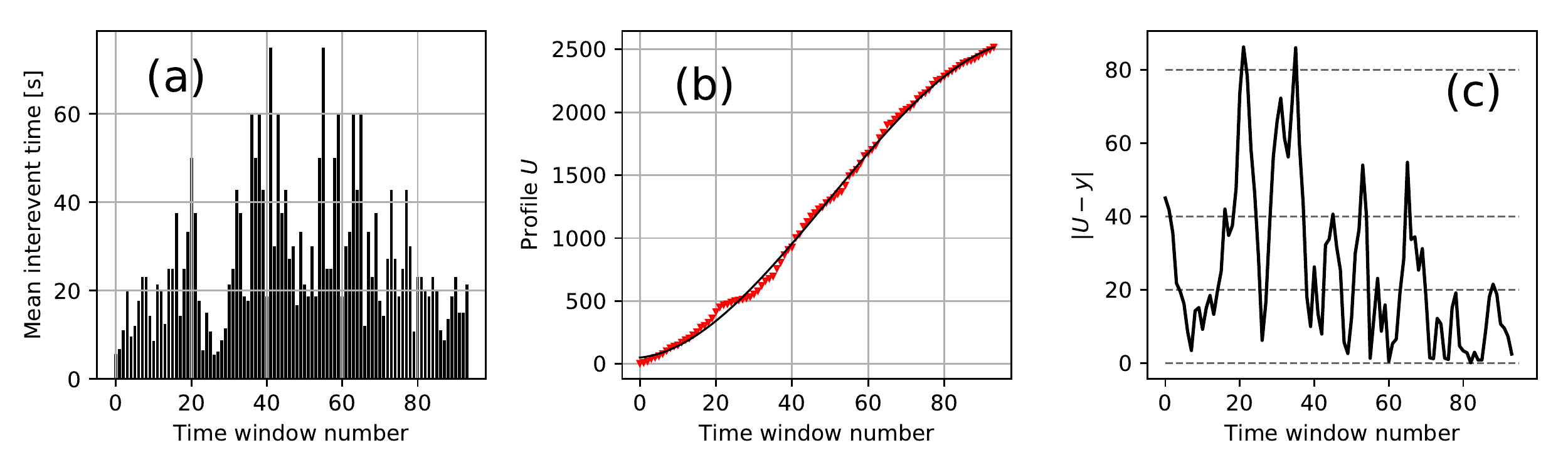}
\caption{Intraday patterns for the 14th January 2011 ($\nu =9$, Friday): typical dependences of significant characteristics vs. time window number $i$. (a) The basic empirical quantity in the form of hierarchical comb-like structure, that is the mean inter-event times $\overline{\Delta t_{i}^{\nu }},~i=1,2,\ldots~,s$ (see Fig.~\ref{figure:scheme} for the definition). (b) The empirical single-day profile $U_{\nu}$ (the monotonically increasing fluctuating red curve consisting of small inverted triangles) defining a formal, directed random walk, together with the fitted smooth thin black curve. This curve represents the best fit by the third-degree polynomial, $y_{\nu }$, well reproducing the inflection point present in the profile. This plot has been supplemented with the plot (c) clearly showing the fluctuating hierarchical structure of the absolute deviations $\mid U_{\nu }(i)-y_{\nu }(i)\mid $ vs. $i$. Horizontal dashed lines roughly mark their amplitude levels.}
\label{figure:profile}
\end{center}
\end{figure*}

Note that 
\begin{eqnarray}
y_{\nu }(i)=\sum_{m=0}^M A_{\nu }^m i^{M-m},\; M\geq 0,
\label{rown:ynu}
\end{eqnarray}
where in all our further considerations we assume $M=3$. This degree of the polynomial enables us to reproduce the inflection point present in the overwhelming majority of empirical daily profiles, $U_{\nu }, \nu =1,2,\ldots $ -- see, for example, the red empirical curve (small red triangles) in plot Fig.~\ref{figure:profile}(b). This is the result of (intraday) \emph{lunch effect} -- cf. plot Fig.~\ref{figure:profile}(a). 
The situation is even more complicated, because we observe two rather than one maximum at lunchtime. We emphasize that the detrending polynomial is fitted to the every-single-day (or $\nu ^{th}$ day) empirical data individually. Therefore, we identify single-day trends, which allows us properly to analyze single-day fluctuations.

For $j=0$ the detrended autocorrelation function (in such a case usually referred to as the detrended self-correlation function) becomes our 'detrended fluctuation function'. Hence, the simplified notation $F^2({\nu },s)\stackrel{\rm def.}{=}F^2(j=0;{\nu },s)$ can be used. 

The single-day profile $U_{\nu }(i)$ for $\nu ^{th}$ day at time window number $i$ is given in Eq. (\ref{rown:Yy}) by the corresponding difference between subsequent multi-day profiles $Y$s. We assume that this difference equals the cumulation of the {\em mean} inter-event times, $\overline{\Delta t_{i'}^{\nu }}$, over time windows (indexed by $i'$) within a single $\nu $-day. The precise definition of this mean, i.e. the mean for $i$th time window, $[i,i+1[,\; 1\leq i\leq s$, (of $i$-independent width $\Delta $) is given by the corresponding time average $\overline{\Delta t_{i}^{\nu }}\stackrel{\rm def.}{=}\frac{1}{n_i^{\nu }}\sum_{l=1}^{n_i^{\nu }}\Delta t_{i,l}^{\nu }$, of inter-event times, $\Delta t_{i,l}^{\nu }$, where $n_i^{\nu }\geq 1$ is the number of subsequent inter-event times belonging to time window number $i$ and trading day number $\nu =1,2,\ldots ,N_d$. Please refer to Fig.~\ref{figure:scheme} for illustration.

The mean inter-event times, $\overline{\Delta t_{i'}^{\nu }}$, are displayed, in the form of a random comb in Fig.~\ref{figure:profile}(a) (in our considerations we deal, in fact, with $\overline{\Delta t _{i'}^{\nu }} \leq \Delta $). Hence, we can write,
\begin{eqnarray}
U_{\nu }(i)&=&Y[(\nu -1)s+i]-Y[(\nu -1)s] \nonumber \\
&=&\sum_{i'=1}^i \overline{\Delta t_{i'}^{\nu }},
\label{rown:Yy}
\end{eqnarray}
where for the first trading day ($\nu =1$) we have $Y[(\nu -1)s=0]=0$ and $U_{\nu =1}(i)=Y[i]=\sum_{i'=1}^i \overline{\Delta t_{i'}^{\nu =1}}$. Therefore, the multi-day profile $Y$ and hence the single-day profile $U$ are based on {\em mean} inter-event times, $\overline{\Delta t_{i'}^{\nu }}$, instead of actual inter-event times, $\Delta t_{i,l}^{\nu }$ themselves. In this sense, these profiles are obtained through a pre-processing stage of coarse-grain type.

Notably, Eq. (\ref{rown:Yy}) makes it possible to determine the multi-day profile recurrently,
\begin{eqnarray}
Y[(\nu -1)s+i]=\sum_{\nu '=1}^{\nu -1}\sum_{i'=1}^s \overline{\Delta t_{i'}^{\nu '}}
+\sum_{i'=1}^i \overline{\Delta t_{i'}^{\nu }},~ \nu \geq 2. \nonumber \\
\label{rown:Ynui}
\end{eqnarray}

Eqs. (\ref{rown:Yy}) and (\ref{rown:Ynui}) can be formally interpreted in terms of the directed (persistent or climbing) random walk -- see the monotonically increasing empirical red broken curve drawn in the plot in Fig.~\ref{figure:profile}(b). If there were such a need (e.g., generalized Hurst exponent would be close to zero) it would be possible to integrate the time series before the procedure, similarly to what is done in the canonical MF-DFA method (cf. Eqs. (7) and (8) in Ref.~\cite{{KZKBHS}}).

Fig.~\ref{figure:profile} is intended to show the intra-day structure of the empirical data. Especially the absolute detrended data shown in the plot in Fig.~\ref{figure:profile}(c) well illustrates the leading but a bit noisy 
hierarchical structure of amplitudes. The levels of this structure are roughly expressed by the series of amplitudes $20\times 2^0,~20\times 2^1,~20\times 2^2$ -- the dashed horizontal lines mark their levels. 

The typical intra-day pattern of single-day mean inter-event times, $\overline{\Delta t_{i}^{\nu }}$, of transactions falling into the $i^{th}$ time window $(i=1,2,\ldots ,s)$ of a given day $(\nu =1,2,\ldots ,N_d)$  for fixed $\nu $ is shown with respect to $i$ in Fig.~\ref{figure:profile}(a). Other plots in Fig.~\ref{figure:profile}(b) and Fig.~\ref{figure:profile}(c) are also plotted vs $i$. Data bursts and explosion of spikes containing hierarchy of singularities are seen well. If such a fluctuation structure were not there, we would not be able to identify multifractality. 

To extract the intra-day structure of fluctuations for a given day we fitted the $\nu$-dependent polynomial $y_{\nu }(i)$ given by Eq. (\ref{rown:ynu}) (see the black curve in plot Fig.~\ref{figure:profile}(b)). This approach is more subtle than the one based on the $\nu $-independent average over statistical ensemble of days $\langle \overline{\Delta t_{i}}\rangle \stackrel{\rm def.}{=}\frac{1}{N_d}\sum_{\nu =1}^{N_d}\overline{\Delta t_{i}^{\nu }}$.

For all the trading days the patterns shown in the plots in Figs.~\ref{figure:profile}(a) and~\ref{figure:profile}(c) look similar, although the corresponding local minima and maxima are somewhat differently distributed having slightly different amplitudes. 

All the plots in Figs.~\ref{figure:profile} and~\ref{figure:Stationar} are prepared for a typical time window of length $\Delta=300$ [sec]. Hence, the daily total number of time windows $s=28~200/300=94$. This is because the duration of a daily stock market session of the Warsaw Stock Exchange, which we consider here, equals $T=7$~[h] $50$ [min] = $28~200$~[sec]. It is worth knowing that the mean number of transactions within a single time window $\Delta=300$ [sec] is about $\langle \langle n\rangle \rangle \stackrel{\rm def.}{=} \frac{1}{N_d}\frac{1}{s}\sum_{\nu =1}^{N_d}\sum_{i=1}^sn_i^{\nu }=20$ as the empirical mean time distance between subsequent transactions approximately equals $\langle \langle \overline{\Delta t}\rangle \rangle \stackrel{\rm def.}{=}\frac{1}{N_d}\frac{1}{s}\sum_{\nu =1}^{N_d}\sum_{i=1}^s\overline{\Delta t_{i}^{\nu }}=15$ [sec] and $\Delta = \langle \langle n\rangle \rangle \langle \langle \overline{\Delta t}\rangle \rangle $. 

Thus, we introduce the averaging over both timescales: intraday $s$ and inter-day $N_d$ ones. The independent detrending of time series segments in the canonical Multifractal Detrended Fluctuation Analysis, we replaced by detrending on a daily timescale combined with averaging over days. In this way, we significantly reduce the statistical error of the results. The triple average introduced above can be a deterrent, but is easily implemented being the repeated simple arithmetic mean.

It is worth noting that the local clusters of spikes around their local maxima are visible in Fig.~\ref{figure:profile}(a) (to a good approximation) four times a day thus not only close to lunchtime. These clusters are separated by the corresponding three interludes of high system activity, where the shortest lengths of inter-event time intervals are present. Therefore, approximately every $100$ min (= $20~\mbox{time windows}~\times \Delta ~(=5~\mbox{min}))$ we see spikes of locally longest lengths. Such a long-term pattern constitutes one of the dominant sources of the volatility clustering effect within the mean inter-event time series. It is a likely result of the existence of long-range autocorrelations between subsequent inter-event times (cf.~\cite{JKTG} and refs. therein). These autocorrelations, we presume to be the source of the true multifractality investigated in this work.

Fig.~\ref{figure:profile} demonstrates the preparatory (pre-processing) stage of our multifractal procedure. Not only this stage differs in essence from the corresponding pre-processing stage of the MF-DFA, but as will become apparent below, other stages also show significant differences. The results presented in Fig.~\ref{figure:profile} (and also in Fig.~\ref{figure:Stationar} in Appendix~\ref{section:appendixB}) constitute the rationale for the subsequent stages of the procedure.

\subsection{Non-monotonic multiscale generalized partition function}\label{section:GPF}

The generalized $q$-dependent or $q$-filtered statistical-mechanic partition function can be defined as usual by the sum,
\begin{eqnarray}
Z_q(s)\stackrel{\rm def.}{=}\sum_{\nu =1}^{N_d}[p(\nu ,s)]^q;
\label{rown:ZqsjD}
\end{eqnarray}
hence, $Z_{q=0}(s)=N_d$ and it is independent of $s$. This independence and division of the full time series into days distinguishes our approach from the multifractal analyzes used so far. It is an important modification which uses the intraday scaling and fluctuations within the statistical ensemble of days.

The probability $p(\nu ,s)$ present in Eq. (\ref{rown:ZqsjD}) specifies the chance of occurrence of a specific fluctuation value for a given day $\nu $ within the scale $s$. This probability, which could be referred to as escort probability as it is escorting the fluctuations, is constructed in the form,
\begin{eqnarray}
p(\nu ,s)=\frac{\left[F^2(\nu ,s)\right]^{1/2}}{\mbox{Norm}(s)}, \nonumber \\
\mbox{Norm}(s)=\sum_{\nu =1}^{N_d}\left[F^2(\nu ,s)\right]^{1/2}
\label{rown:pjnus}
\end{eqnarray}
that is, it is based on the fluctuation function defined by Eq. (\ref{rown:F2snj}) for $j=0$. Hence, the mean value, $\langle p(s)\rangle = 1/N_d \sum_{\nu =1}^{N_d}p(\nu ,s) = 1/N_d$, is fixed (as a result of normalization). An even more refined approach based on a $q$-zooming escort probability has been established in Ref.~\cite{PJTA}.

We introduce the scaling hypothesis in the usual approximate form~\cite{KZKBHS}, which we verify in Fig.~\ref{figure:Fqj0},
\begin{eqnarray}
\sum_{\nu =1}^{N_d}\left[F^2(\nu,s)\right]^{q/2}\approx N_dA_q s^{qh(q)},
\label{rown:scalZ}
\end{eqnarray}
where the prefactor $A_q$, and the generalized Hurst exponent $h(q)$ are $s$-independent quantities. Besides, by putting $q=0$, one directly obtains the constrain $A_{q=0}\approx 1$ from the scaling hypothesis (\ref{rown:scalZ}).

Note, that Eq. (\ref{rown:scalZ}) allows the presentation of $\mbox{Norm}(s)$ (given by the second equality in Eq. (\ref{rown:pjnus})) in the form,
\begin{eqnarray}
\mbox{Norm}(s)=\sum_{\nu =1}^{N_d}\left[F^2(\nu,s)\right]^{1/2} 
\approx N_dA_{q=1} s^{h(q=1)}, \nonumber \\
\label{rown:scaNN}
\end{eqnarray}
by putting $q=1$.

Indeed, Appendix~\ref{section:appendixC} uses Eqs. (\ref{rown:ZqsjD}) -- (\ref{rown:scaNN}) for the presentation of significant properties of $Z_q(s)$ and related quntities. The point is that with Eqs. (\ref{rown:ZqsjD}) -- (\ref{rown:scaNN}), we obtain a scaling exponent $\tau (q)$ defined with Eq. (\ref{rown:tauq}). Alongside the generalized Hurst exponent, this is another pillar of multifractality. We consider exponents $h(q)$ and $\tau (q)$ as well as other multifractality characteristics below.

\subsection{Legendre-Fenchel transformation and multi-branched multifractality}\label{section:L-F_transform}

In this section, we carry out our multi-branched multifractal analysis on the example of the time series of inter-event times.

Equation (\ref{rown:scalZ}) can be written in an equivalent form more convenient for application to empirical data,
\begin{eqnarray}
\ln {\cal F}_q(s) \approx h(q)\ln s + B(q),
\label{rown:calFqs}
\end{eqnarray}
where $q$-dispersive 
\begin{eqnarray}
{\cal F}_q(s)\stackrel{\rm def.}{=}\left\{N_d^{-1}\sum_{\nu =1}^{N_d}\left[F^2(\nu,s)\right]^{q/2}\right\}^{1/q} 
\label{rown:calFqs2}
\end{eqnarray}
and $B(q)\stackrel{\rm def.}{=}q^{-1}\ln A_q$.

\begin{figure*}
\begin{center}
\includegraphics[scale=1.50,angle=0,clip]{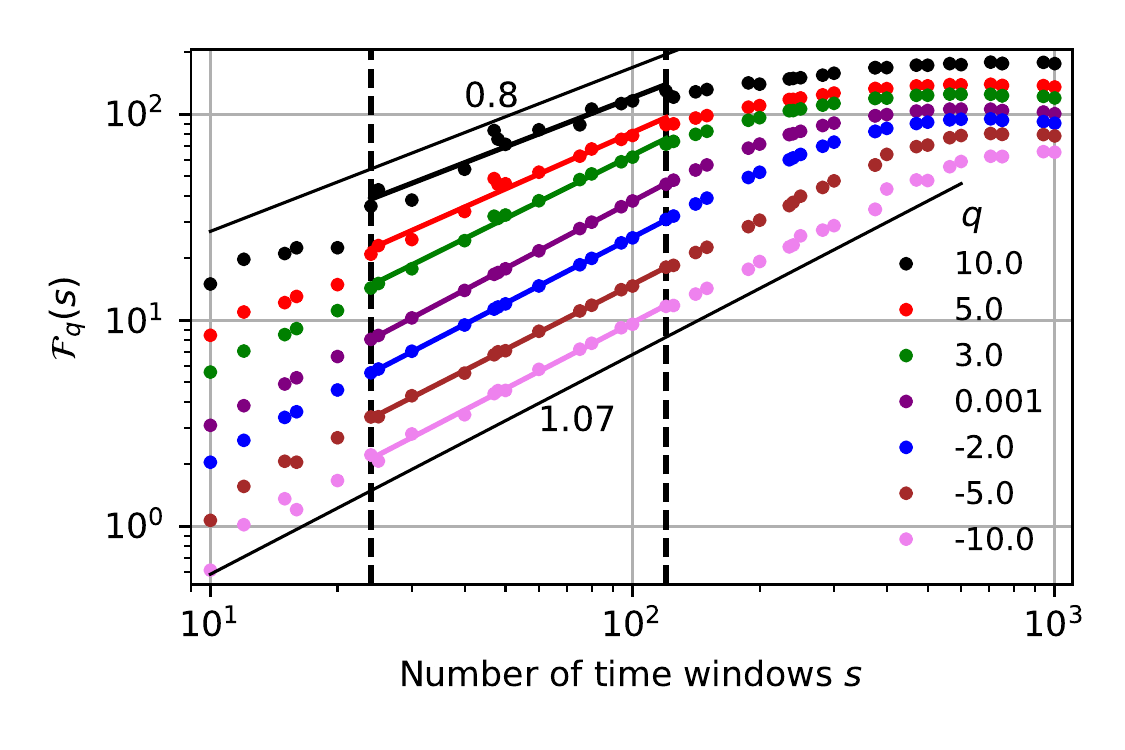}
\caption{Plots of the function ${\cal F}_q (s)$ (defined by Eq. (\ref{rown:calFqs2})) vs. $s$ within the log-log scale for the extended range of $-10\leq q\leq 10$. Vertical dashed lines define the common region of the best fit of all the straight lines -- the first one at $s=24$ or $19~\mbox{min}~35~\mbox{s}$ and the second one at $s=120$ or $3~\mbox{min}~55~\mbox{s}$. The upper and lower sloped thin black straight lines define the boundaries (here, between 0.80 and 1.07, still far beyond statistical errors) in which the slope of the fitted lines is included. It should be emphasized that between q = -5 (straight brown line fitted to the empirical data) and q = 5 (the fitted straight red line) lie straight lines (purple and green) fitted to the empirical data, the slope of which is slightly greater than the slope of the two mentioned above lines. This indicates non-monotonic dependence  on $q$ of the slope considered.}
\label{figure:Fqj0}
\end{center}
\end{figure*}

\begin{figure*}
\begin{center}
\includegraphics[scale=0.48,angle=0,clip]{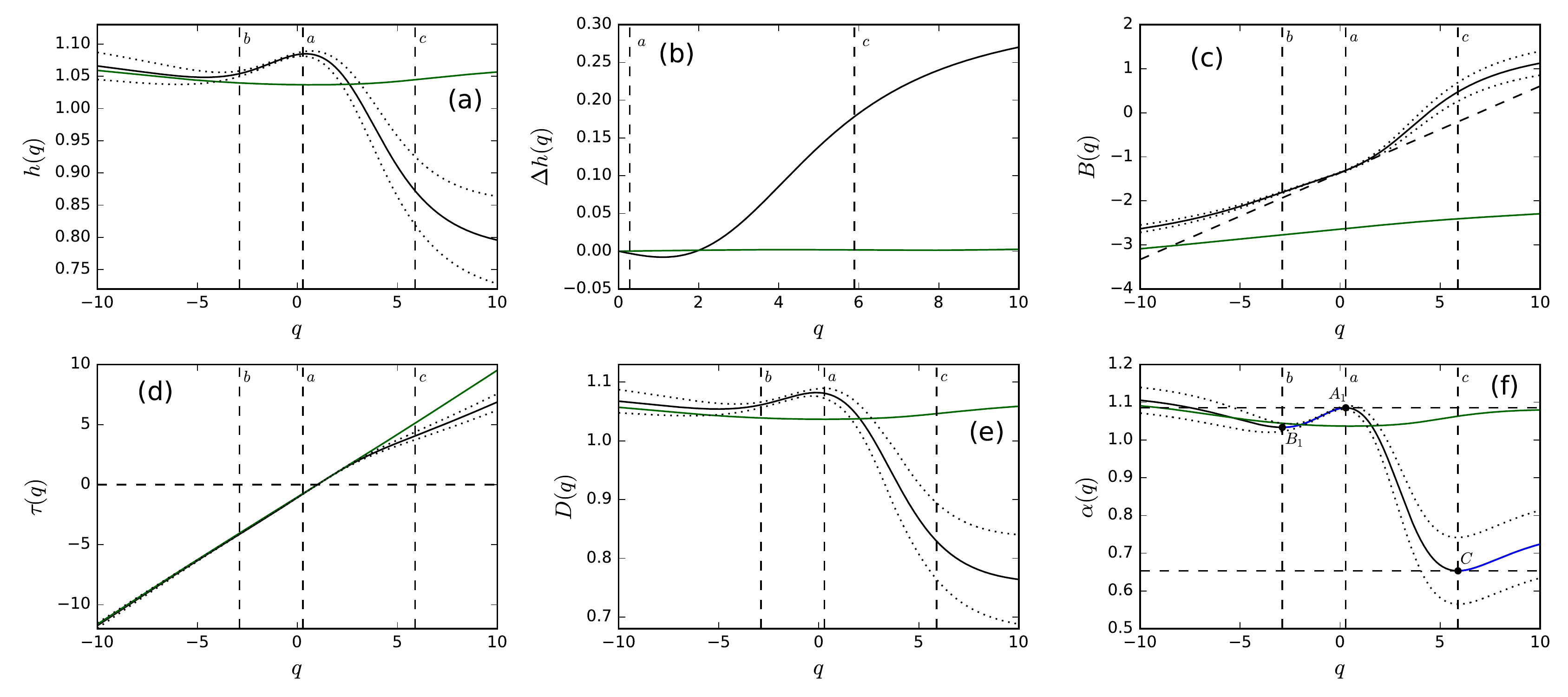}
s\caption{We present the $q$-dependence of key empirical characteristics of multifractality. The nonlinear dependence of these characteristics on $q$ is visible in all the plots (black curves, while dotted curves designate corresponding one-standard-deviation error bounds; in addition, fragments $[B_1,A_1]$ and that starting from point $C$ of the curve $\alpha (q)$ are blue). Plots (a) -- (f) present dependence on $q$ of the generalized Hurst exponent $h(q)$, its spread $\Delta h(q)=h(-q)-h(q)$, prefactor $B(q)$ related to reduced R\'enyi information, R\'enyi scaling exponent $\tau (q)$, R\'{e}nyi dimensions $D(q)$, and the coarse H\"older exponent $\alpha (q)$, respectively. The straight vertical dashed lines $a$ and $c$ define the range of $q$-support located between the absolute maximum $A_1$ and absolute minimum $C$ of curve $\alpha (q)$ shown in plot (f) vs $q$. A similar line $b$ indicates the location of the second minimum $B_1$ of the curve $\alpha (q)$. Analogous lines we also applied to the remaining plots. The tangent dashed straight line visible in the plot (c) we fitted to the linear section of the curve $B(q)$ vs. $q$. The additional thin (green) solid curves present in all the plots we obtained from the time series (of the same size as the empirical ones) generated from the Poisson distribution. Their variations are negligible. Therefore, the influence of finite size effects on time series of the size considered is negligible.}
\label{figure:Hurstqj}
\end{center}
\end{figure*}

Using the dependence of ${\cal F}_q(s)$ on the scale $s$ (see Fig.~\ref{figure:Fqj0} for details) 
for values of $q$ from its extended range (that is $-10\leq q\leq 10$), we have determined all the essential multifractal characteristics. These are: the generalized Hurst exponent $h(q)$, its spread $\Delta h(q)=h(-q)-h(q)$, significant prefactor $B(q)$ present in Eq. (\ref{rown:calFqs}) related to reduced R\'enyi information, related signatures of multifractality such as R\'enyi scaling exponent $\tau (q)$, R\'enyi dimensions $D(q)$, and the coarse H\"older exponent $\alpha (q)$ (see plots in Fig.~\ref{figure:Hurstqj} for details). In the range of $q$ considered, we observed non-monotonic behavior of the slopes of the respective curves vs. $q$ which is crucial for this work.

The fitting in the narrower area of $10\leq s\leq 120$ is characterized by a small statistical error that makes it possible to observe the non-monotonicity mentioned, in which the statistical error is substantially higher than the statistical error of fitting. This is further considered in the text below.

Our analysis in the range $-10\leq q \leq 10$ shows not only the non-linear behavior of the scaling exponent $\tau $ vs. $q$ (and dependencies of the related characteristics), but also what the asymptotes of the scaling exponent $\tau (q)$ look like. This is visualized in the plot in Fig.~\ref{figure:tauqcomp} (the range of the variable $q$ is a bit narrower $-5\leq q\leq 7$ here, to better show the non-linear relationship in the central part). However, only the finer analysis shown in Fig.~\ref{figure:Hurstqj} shows the correct range of the variable $q$. This is a more subtle (local) analysis because it operates using functions based on the derivative of $\tau (q)$ relative to the variable $q$. It can be seen, e.g. in Fig.~\ref{figure:Hurstqj}(f), that the extended range of the variable $q$ selected above is appropriate due to the presence of non-monotonicity. It is this extended range of $q$ that distinguishes our analysis from the standard MF-DFA.
\begin{figure*}
\begin{center}
\includegraphics[scale=0.48,angle=0,clip]{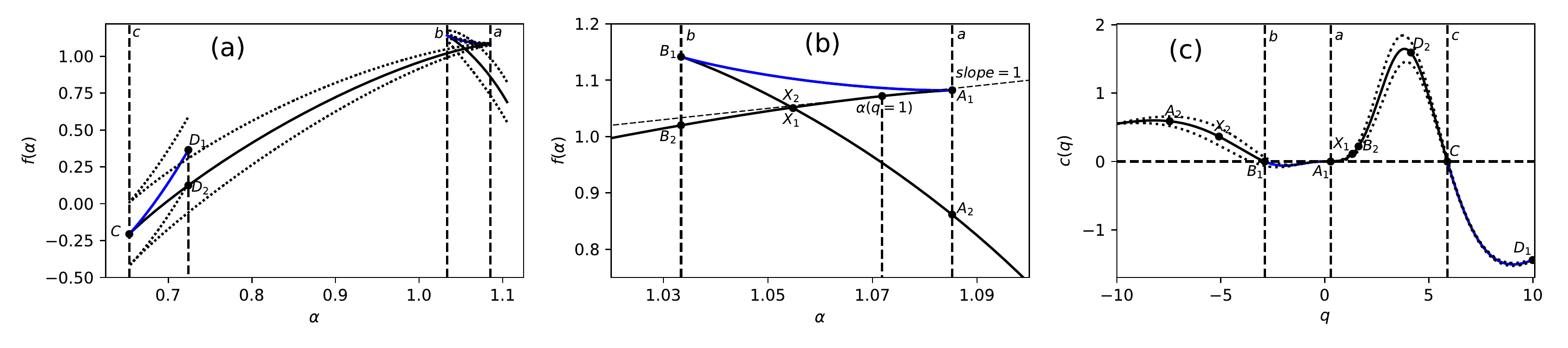}
\caption{Two plots (a) and (b) (where (b) is an enlarged part of plot (a)) show complementary views of the spectrum of dimensions $f(\alpha )$ (given by Eq. (\ref{rown:fqalpha})) vs $\alpha $. In these plots the main branch of the spectrum $f(\alpha )$ is an increasing black curve. Plot (c) shows the specific heat $c(q)$ (given by Eq. (\ref{rown:cq})) vs $q$, used clearly to define thermal stable and unstable phases. The unstable phase area is defined by the negative specific heat. 
The shorter vertical dashed straight line displayed in the plot (b) at the point $\alpha (q = 1)$, determines the position {of the single contact point} of the tangent dashed straight line with a slope of $1.0$ for the main branch. It is a verification of the contact property of the L-F transform. Other dashed lines are described in the main text, while a pair of points $X_1,X_2$ and $D_1,D_2$ are further addressed in the context of Fig.~\ref{figure:Singularf} below. The dotted curves designate corresponding one-standard-deviation error bounds.}
\label{figure:cqvsq}
\end{center}
\end{figure*}

Let us emphasize that all these quantities and their standard deviations, together with the multifractal spectrum $f(\alpha )$ considered below, were obtained from empirical data using Eq. (\ref{rown:calFqs}) for direct estimation. Other quantities were obtained indirectly.

By taking the scaling exponent $\tau (q)$ from Eq. (\ref{rown:tauq}), the coarse H\"older exponent $\alpha (q)$ and multi-branched multifractal spectrum $f(\alpha )$ can be found. We define below the Legendre-Fenchel (LF) transformation which we use instead of the standard Legendre transform. Although formally both look the same, LF transform allows a multi-branch solution (see Appendix~\ref{section:appendixG} for additional information). We have
\begin{eqnarray}
\alpha (q)&\stackrel{\rm def.}{=}&\frac{d\tau (q)}{dq}, \nonumber \\
f(\alpha )&\stackrel{\rm def.}{=}&q\alpha (q)-\tau (q), 
\label{rown:fqalpha}
\end{eqnarray}
hence,
\begin{eqnarray}
q&=&\frac{df(\alpha (q))}{d\alpha }~\mbox{and}~f=-\frac{d\left(\tau (q)/q\right)}{d\left(1/q\right)},
\label{rown:dfqalpha}
\end{eqnarray}
where $\alpha $ is a local dimension (singularity or coarse H\"older exponent -- its $q$-dependence is shown in Fig.~\ref{figure:Hurstqj}(f)), while $f(\alpha)$ is its distribution shown in the plots in Fig.~\ref{figure:cqvsq}(a) and Fig.~\ref{figure:cqvsq}(b). As usual, for a monofractal structure the scaling exponent $\tau (q)$ is a linear function of $q$, while for a multifractal the dependence is non-linear.

We are considering the multi-branch function only of the type shown in Fig.~\ref{figure:cqvsq}(a) and Fig.~\ref{figure:cqvsq}(b). Specifically, we consider a function consisting of a concave main branch -- between the points $A_1$ and $C$ -- and smoothly attached lateral side branches. The term `smoothly' here means that at the turning points the derivatives calculated along the separate branches are equal at the points where the branches meet. That is, our multi-branch (multi-valued) function (mapping) is differentiable at any point. 

Thus, formulated above is the operational definition of the multi-branch spectrum of dimensions. Formally, {\em multi-branch spectrum} is defined as a spectrum for which the second derivative of $f(\alpha)$ with respect to $\alpha$ is discontinuous. A {\em branch} of the spectrum is, therefore, defined as a segment of the spectrum $f(\alpha)$ for which the second derivative with respect to $\alpha$ is continuous. Each branch is, therefore, bounded by the points of discontinuity of $d^2f/d\alpha^2$. Further discussion of the role of the discontinuities in  $d^2f/d\alpha^2$ is carried in the context of phase transition phenomena in Sec.~\ref{section:fsphasetr}.

From Eq. (\ref{rown:tauq}) and the {first} equality in Eq. (\ref{rown:fqalpha}) we obtain the expression, 
\begin{eqnarray}
\alpha (q)=h(q)+q\frac{dh(q)}{dq}
\label{rown:alph}
\end{eqnarray}
{
Hence, at $q=0$ and $q=q_{extr}^1,q_{extr}^2$ we obtain $\alpha (q)=h(q)$, where $q_{extr}^j,~j=1,2,$ defines $q$-positions of local extrema of $h(q)$ function (which are not marked in Fig.~\ref{figure:Hurstqj}(a) because they are clearly visible). However, these do not imply the extrema of $\alpha (q)$ function (see Appendix~\ref{section:appendixD} for details).

Figs.~\ref{figure:Hurstqj}(a),~\ref{figure:Hurstqj}(f),~\ref{figure:cqvsq}(a), and~\ref{figure:cqvsq}(b) form the graphical basis of this work.}

It is worth emphasizing that aggregating events into time intervals of the same length ($\Delta $, see Fig.~\ref{figure:scheme} for details) may influence the analysis. Namely, if the intervals are too short, too many of them will be empty. On the other hand, if the intervals are too long, aggregation of too many points may lead to the loss of information on the time structure of the process.

The analysis shown in Fig.~\ref{figure:Fqj0} proposes a solution to this problem. This is based on selecting an appropriate range of $\Delta $ in which the scaling effect is observed. Here, for ${\cal F}_q(s)$ vs $s=T/\Delta $, where $T=7~\mbox{h}~50~\mbox{min}$ or $470~\mbox{min}$ and for the all considered values of $-10\leq q\leq 10$ we have a common range $3~\mbox{min}~55~\mbox{sec}\leq \Delta \leq 19~\mbox{min}~35~\mbox{sec}$. 
For this range of $s$, the measure $\chi ^2$ per degree of freedom reaches the smallest value. This quantity is only marginally larger when the left border of $s$ is extended to $s=10$ or $\Delta =47~\mbox{min}$, while the right one is still kept at $s=120$ or $\Delta =3~\mbox{min}~8~\mbox{sec}$. However, it is then burdened with a larger fitting error which we wanted to avoid. In addition, we wanted the fit within the $s$ range to be common to all considered curves.

We could, of course, vary the upper edge of the fit intervals (e.g., for $q=-5$, it could be $s\approx 500$ instead of $120$), but this would also increase the fitting error. The natural limits of the fitting range from $s_{min}\approx 1$ to $s_{max}\approx 1000$ and exceeding these limits would pose problems. Below $s_{min}\approx 1$ we would have a situation where the mean masks any variation (since then $\Delta \approx T$). Above $s_{max}\approx 1000$, there are time intervals of width $\Delta $ that do not contain any transactions. In this way, we determine the natural boundaries of the scaling area.

It must be clearly stated that due to the non-monotonic dependence of the generalized Hurst exponent $h(q)$ versus $q$, the spectrum of dimensions $f(\alpha )$ is a multi-branched function of the H\"older $\alpha $ exponent (see plots Fig.~\ref{figure:cqvsq}(a) and Fig.~\ref{figure:cqvsq}(b) for details).

Recall that the Legendre transformation only deals with monotonous functions $h(q)$. From this point of view, Eqs. (\ref{rown:fqalpha}) and (\ref{rown:dfqalpha}), although formally identical to the Legendre transform, are its generalization. The Legendre transform is limited here only to the main branch of the spectrum $f$ defined by its contact relations: 
\begin{itemize}
\item[(i)] $f(\alpha (q=1)) = \alpha (q=1)$
\item[(ii)] $\frac{df}{d\alpha (q)}|_{\alpha (q=1)}=1$.
\end{itemize}
The inset plot present in Fig.~\ref{figure:cqvsq}{(b)} illustrates this contact character. This is emphasized by a dashed straight line with directional coefficient (slope) of $1.0$ tangent to spectra of singularities at the point $\left[\alpha (q=1),f(\alpha (q=1))\right]$. Breaking the contact character of the Legendre transformation results in the wrong location of the spectrum of singularities. 

Put more generally, the contact relations given above (for $q=1$) provide an unambiguous location of the full multi-branched spectrum of dimensions obtained using the Legendre-Fenchel transformation. Our multi-branched multifractal contains a single contact point which implies that we are dealing with a  multi-branched multifractal. Fig.~\ref{figure:cqvsq}(a) and Fig.~\ref{figure:cqvsq}(b) illustrate an important result, namely a necessary (not sufficient) requirement for finding true multifractality in empirical time series.

Finally, thanks to the above consideration, we can clarify the key term `multi-branched multifractality' with a dominant left-sided branch. We are dealing with this type of multifractality if the location of the main branch of the spectrum of dimensions is fully determined by $q > 0$. The main branch of the spectrum is the one that meets the condition of contact. This is visible in Fig.~\ref{figure:Hurstqj}(f), depicting the relationship between $\alpha $ and~$q$. 

In Fig.~\ref{figure:cqvsq}(b), we mark with $\alpha(q = 1)$ and with a vertical dashed line, the location of the contact point with straight line -- the dashed line having a slope equal to 1. The right border of this main branch is at point $A_1$, whose coordinate $q$ is slightly greater than zero, and the left border is at point C (see again Fig.~\ref{figure:Hurstqj}(f) for details).

\section{First and second order phase transitions}\label{section:fsphasetr}

The multi-branched multifractality obtained by us has a useful, fundamental thermodynamical interpretation. Studying a thermodynamical interpretation is a standard way to analyze the properties of multifractals. The indicator used for thermodynamical classification is the specific heat of the multifractal structure~\cite{KKPM} (and refs. therein).

From Eq. (\ref{rown:fqalpha}) one can obtain a useful expression for the specific heat of the multifractal structure in the form,
\begin{eqnarray}
c(q) = \frac{d\alpha (q)}{d(1/q)} = -q^2\frac{d\alpha (q)}{dq};
\label{rown:cq}
\end{eqnarray}
its $q$-dependence is shown in Fig.~\ref{figure:cqvsq}(c). 

Only two regions are visible in which the system is thermally stable, i.e. fulfilling inequality $c(q)\geq 0~(\mbox{or}~\frac{d\alpha (q)}{dq)}\leq 0)$. The first of them is located between dashed vertical straight lines $a$ and $c$ or points $A_1$ and $C$ (the same as shown in Fig.~\ref{figure:Hurstqj}f). Thus, we defined the $q$-range of the main branch of the spectrum of dimensions. We show this branch in Fig.~\ref{figure:cqvsq}(a) and Fig.~\ref{figure:cqvsq}(b) (the monotonically increasing black curves in both plots). 

The second region is limited to the range of $q$ preceding vertical dashed straight line $b$ presented in plots Fig.~\ref{figure:Hurstqj}(a), (c)--(f) or point $B_1$ in plot Fig.~\ref{figure:Hurstqj}(f). In this way we get $q$-support of the side-branch spectrum of dimensions (the decreasing black curve in plot Fig.~\ref{figure:cqvsq}(b) starting at point $B_1$ and passing through points $X_1, X_2$, and $A_2$). 

\begin{figure*}
\begin{center}
\includegraphics[scale=0.80,angle=0,clip]{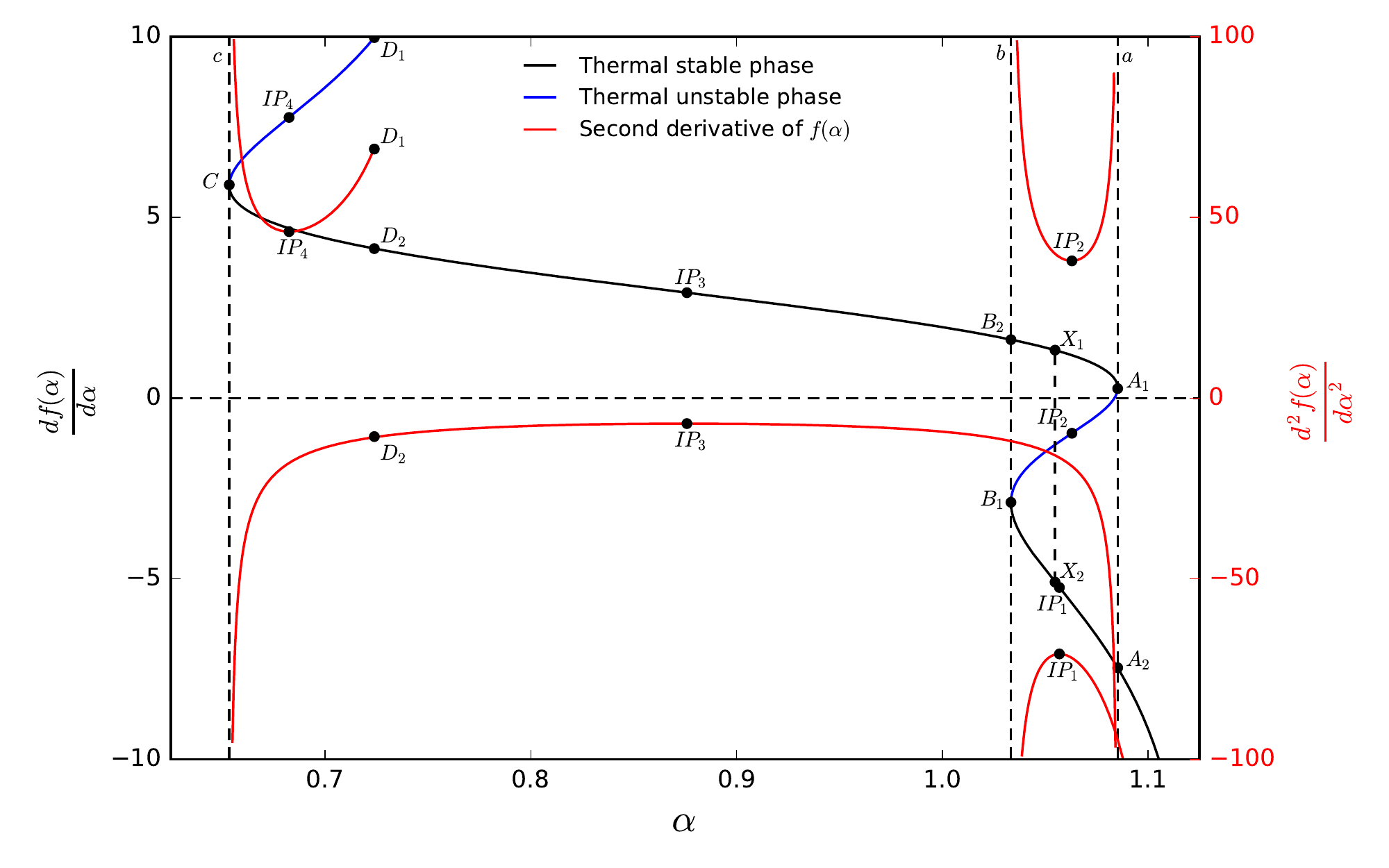}
\caption{The schematic illustration of the modified Ehrenfest or Mandelbrot classification of phase transitions~\cite{GrJae}. The first (black and blue fragments of a single solid curve) and second (four separated red solid curves) order derivatives of $f$ over $\alpha $ showing three two-branched second-order singularities of $f$ vs. $\alpha $. Three dashed vertical straight lines (vertical asymptotics) marked by $c,b$, and $a$ located at $\alpha $ coordinates of these singularities that is defined by points $C$, $B_1$, and $A_1$.  In the main text of this section, we provide a detailed discussion of this figure.}
\label{figure:Singularf}
\end{center}
\end{figure*}

Moreover, between points $X_1$ and $X_2$ located in thermally stable phases, the first-order phase transition occurs. It is considered below in the context of Fig.~\ref{figure:Singularf}. 

A peculiar characteristic of our multifractal is the presence of a negative spectra of dimensions, in the vicinity of the turning point $C$, in Fig.~\ref{figure:cqvsq}(a), which could be justified by the appearance of events that occur exceptionally rarely (see~\cite{JPV} for some suggestions).

We deal with thermally unstable phases for the opposite case $c(q)< 0~(\mbox{or}~\frac{d\alpha (q)}{dq)}> 0)$. They range between turning points $B_1,A_1$ and after the point $C$, presented in Fig.~\ref{figure:Hurstqj}(f) and Fig.~\ref{figure:cqvsq}(c). In Fig.~\ref{figure:cqvsq}(b) this range of $q$ is clearly visible. It is the $q$-support of the bifurcating branch of our multi-branched multifractal (the solid blue curve stretching between points $B_1$ and $A_1$; in Fig.~\ref{figure:cqvsq}(a) the highest placed short blue curve represents it stretched between dashed vertical lines $a$ and $b$). In points $B_1, A_1$, and $C$, there are phase transitions of the second-order between thermally stable and unstable phases, which is consistent with specific heat vanishing there.  This is discussed below in the context of Fig.~\ref{figure:Singularf} (together with a description of the role of points $D_1$ and $D_2$). 

To prove the statements above concerning the order of phase transitions, we study the behavior of the first, $df/d\alpha $, and second, $d^2f/d\alpha ^2$, derivatives vs $\alpha $, based on the results presented in Fig.~\ref{figure:Hurstqj}(f). Using the Taylor expansion of $\alpha (q)$ function in the vicinity of its local extremes we obtain,
\begin{eqnarray}
\alpha (q) \approx \alpha (q_{extr})+\frac{1}{2}(q-q_{extr})^2\frac{d^2\alpha }{dq^2}\mid _{q=q_{extr}}, \nonumber \\
\label{rown:expans}
\end{eqnarray}
where $q_{extr}$ is a $q$-position of the local extremum or turning point of $\alpha (q)$ function. There are three such local extrema: one maximum  $A_1$ and two minima $B_1,C$.

Inverting Eq. (\ref{rown:expans}) and using the first equation in (\ref{rown:dfqalpha}), after simple algebraic calculations, we obtain useful two-branched formulas,
\begin{eqnarray}
\frac{df}{d\alpha }\approx \pm \sqrt{2\mid \frac{\alpha -\alpha _s}{\ddot{\alpha _s}}\mid }+q_{extr}, \nonumber \\
\frac{d^2f}{d\alpha ^2}\approx \pm \frac{1}{\sqrt{2\mid \ddot{\alpha _s}\mid }}\frac{1}{\sqrt{\mid \alpha - \alpha _s\mid }},
\label{rown:final12}
\end{eqnarray}
where we use the abbreviated notation: $\alpha _s =\alpha (q_{extr})$ and $\ddot{\alpha _s} = \frac{d^2 f}{d\alpha ^2}\mid _{q=q_{extr}}$. The spectrum of dimensions $f$ has singularities of the second order at its turning points (see Fig.~\ref{figure:cqvsq}(a), Fig.~\ref{figure:cqvsq}(b), and Fig.~\ref{figure:Singularf} for illustration).

Moreover, by substituting the expansion given by Eq. (\ref{rown:expans}) to Eq. (\ref{rown:cq}), we obtain 
\begin{eqnarray}
c(q)\approx -q^2(q-q_{extr})\ddot{\alpha _s},
\label{rown:cqlin}
\end{eqnarray}
i.e., it linearly vanishes at the turning points, which can be considered to be spinodal decomposition points (see Fig.~\ref{figure:Singularf} for details).

Fig.~\ref{figure:Singularf} shows the behavior of the first ($df/d\alpha $) and second ($d^2f/d\alpha ^2$) order derivatives of spectrum of dimensions ($f$) versus H\"older exponent ($\alpha $). In combination with the plots (a) and (b) in Fig.~\ref{figure:cqvsq}, this allows us to classify phase transitions at points $A_1$, $B_1$, and $C$ and at a point marked twice by $X_1,X_2$.

The modified Ehrenfest or Mandelbrot classification of phase transitions which we consider here, is based on the spectrum of dimensions $f$, which we treat as the analogon of entropy~\cite{BeSch,KKPM}. Therefore, the classification suggested by Mandelbrot~\cite{ManEvHa,ManEv} can be considered as {\em merely} inspired by Ehrenfest's which uses chemical potential and not entropy. Although both quantities are functions of the thermodynamic state of the system, the Mandelbrot classification is by one order of magnitude lower than the Ehrenfest classification -- this is because entropy is a partial derivative of the chemical potential. 

Although both quantities are functions of the thermodynamic state of the system, they are not equivalent. In the case of the Ehrenfest classification the order of phase transition is determined by the highest possible order of the derivative of the chemical potential versus temperature. In the Mandelbrot classification this order of phase transition is similarly determined by the highest possible order of the derivative, however, in this case the derivative of entropy with respect to temperature -- which in the case of multifractality is played by $1/q$. The use of the Mandelbrot classification in the case of multifractals is more convenient from a technical point of view since entropy is obtained directly from the LF transformation given by Eq. (\ref{rown:fqalpha}) in contrast to the chemical potential.

Both $f$ and $df/d\alpha $ are continuous functions of $\alpha $ as opposed to $d^2f/d\alpha ^2$ (see Figs.~\ref{figure:cqvsq} and~\ref{figure:Singularf} for details) -- both derivatives are calculated numerically as the numerical dependence of $f(\alpha )$ from $\alpha $ is known (cf. Fig.~\ref{figure:cqvsq}). All these functions are multi-branched but only the second order derivative consists of separated branches (cf. red curves in Fig.~\ref{figure:Singularf}). The discontinuity of $d^2f/d\alpha^2$  serves as the formal definition of multi-branching effect as introduced in Sec~\ref{section:L-F_transform}.

All except one, these separated branches diverge asymptotically to $\pm \infty $ in turning points $A_1, B_1$, and $C$ (the corresponding vertical asymptotics denoted by the dashed lines $a,~b,~c$ are presented in Fig.~\ref{figure:Singularf} and also in Figs.~\ref{figure:Hurstqj} and~\ref{figure:cqvsq}). These asymptotic divergences happen according to the power-law with an exponent equal to $-1/2$ -- see the second equality in Eq. (\ref{rown:final12}). Therefore, in these points, there are identical phase transitions of the second order according to our classification, i.e. belonging to the same universality class) -- this is confirmed by the behavior of specific heat given by Eq. (\ref{rown:cqlin}). That is, at the points of the second-order phase transition, specific heat, susceptibilities and other appropriate order parameters either diverge (obeying a non-trivial scaling law), or go to zero -- which case happens in our situation (cf. Fig.~\ref{figure:cqvsq}(c)).

The main branch of the derivative $df/d\alpha $ is represented by the black curve $(C,D_2,B_2,X_1,A_1)$ containing the inflection point $IP_3$ (the corresponding curve in Fig.~\ref{figure:Hurstqj}(f) has a less detailed description). The corresponding second order derivative $d^2f/d\alpha ^2$ (red curve containing a replica of point $D_2$ and inflection point $IP_3$ -- where 'replica' identifies inflection point of the first derivative) diverges to $-\infty $ at asymptotics $c$ and $a$. Therefore, this curve is singular at turning points: its left arm at $\alpha $ coordinate of point $C$ and the right one at $\alpha $ coordinate of point $A_1$. 

The other three separated singular curves (also in red) are associated with three side branches of the first-order derivative $df/\alpha $. The most upper one (located in the left part of the figure, ending at a replica of point $D_1$), has its local minimum at a replica of the inflection point $IP_4$. This curve is bound to side branch $(C,D_1)$ (short blue curve) of the first derivative, containing the inflection point $IP_4$. This branch is thermally unstable (see the plot in Fig.~\ref{figure:cqvsq}(c) for details) as heat capacity is negative. The upper curve (placed at the right part of the figure), having its local minimum at a replica of the inflection point $IP_2$ (also marked by $IP_2$), is bound to branch $(A_1,IP_2,B_1)$ (short blue curve) of the first-order derivative. We consider points $A_1$ and $B_1$ as spinodal decomposition points -- there is a thermally unstable territory between them (see plot in Fig.~\ref{figure:cqvsq}(c) for details again). The third singular solid curve, having its local maximum at a replica of the inflection point $IP_1$ (also denoted by $IP_1$), is bound to branch $(B_1,IP_1,A_2)$. Its left branch has asymptotics at point $B_1$, while its right branch has no asymptotics at the point $A_2$.

Of course, all the branches of the first derivative we associate with the corresponding branches of the spectrum of dimensions, $f$ vs. $\alpha $, clearly shown in plots Fig.~\ref{figure:cqvsq}(a) and Fig.~\ref{figure:cqvsq}(b).

Let us note that short black $(X_1,A_1)$ and $(B_1,X_2)$ curves define the thermally metastable phases, while the short blue curve $(A_1,B_1)$ defines the unstable mixture of phases -- i.e. phases that are described by the black $B_2,X_1,A_1$ and $B_1,X_2,A_2$ curves. If the system is found to be in a mixture phase, it will spontaneously evolve towards a state, which favors either higher fluctuations (defined by $q$ larger than that for point $A_1$) or smaller fluctuations (defined by $q$ lower than that for point $B_1$). The probability of choosing one of these two options depends on how closely the state of the system is located to the edge of the phase.

For the unstable phase defined by the short blue curve $(C,D_1)$ containing the inflection point $IP_4$,  we develop a simplified interpretation. It is because we did not locate it between two metastable phases, although $d^2f/d\alpha ^2$ diverges at transition point $C$ in the same way as at points $A_1$ and $B_1$. We can only say that the system left alone in this phase will spontaneously evolve into a stable phase. 

In Fig.~\ref{figure:Singularf} we present the first-order (discontinuous) phase transition between single phases by the short vertical dashed line connecting $X_1$ and $X_2$ points. Notably, in Fig.~\ref{figure:cqvsq}(b) this first-order phase transition point is marked twice by $X_1$ and $X_2$ -- this defines a single point of branch intersection. 

\section{Concluding remarks}\label{section:disconc}

This work concerns problems of long-term dependence, long-term memory, and long-term/range correlations in time series~\cite{AFB}. By using the Legendre-Fenchel (or generalized Legendre) transform we have examined multi-branched, non-spurious multifractal properties of time series of inter-event times. We have chosen inter-event times for our research because they are a crucial measure of the dynamical activity for many systems (not necessarily complex) -- the research into which is only at the initial stage. The relationships between inter-event times form the foundation of other dynamic relationships occurring in evolving systems, for example, between inter-event times and stochastic process. The time series of inter-event times gives insight into the non-monotonic behavior of the generalized Hurst exponent -- the principal subject of our study.

Our research focuses on the search for multifractality because it is the most general, characterization of time series as of yet. It enables the study of the universal properties from an extended point of view, allowing their classification by using their singularity spectra or spectra of dimensions. However, deriving a microscopic model from the knowledge of the multifractal structure of series of inter-event times is still under consideration. We proposed in Ref.~\cite{KKPM} a step in this direction, where the surrogate model was the Continuous-Time Random Walk with a waiting-time distribution weighted by stretched exponential, i.e. defined by some superstatistics.  It is an approach sufficient to describe multifractality generated by a broadened distribution, but in the case of multifractality caused by long-term auto-correlations of inter-event times, it is still a significant challenge.

\begin{figure}
\begin{center}
\includegraphics[scale=0.70,angle=0,clip]{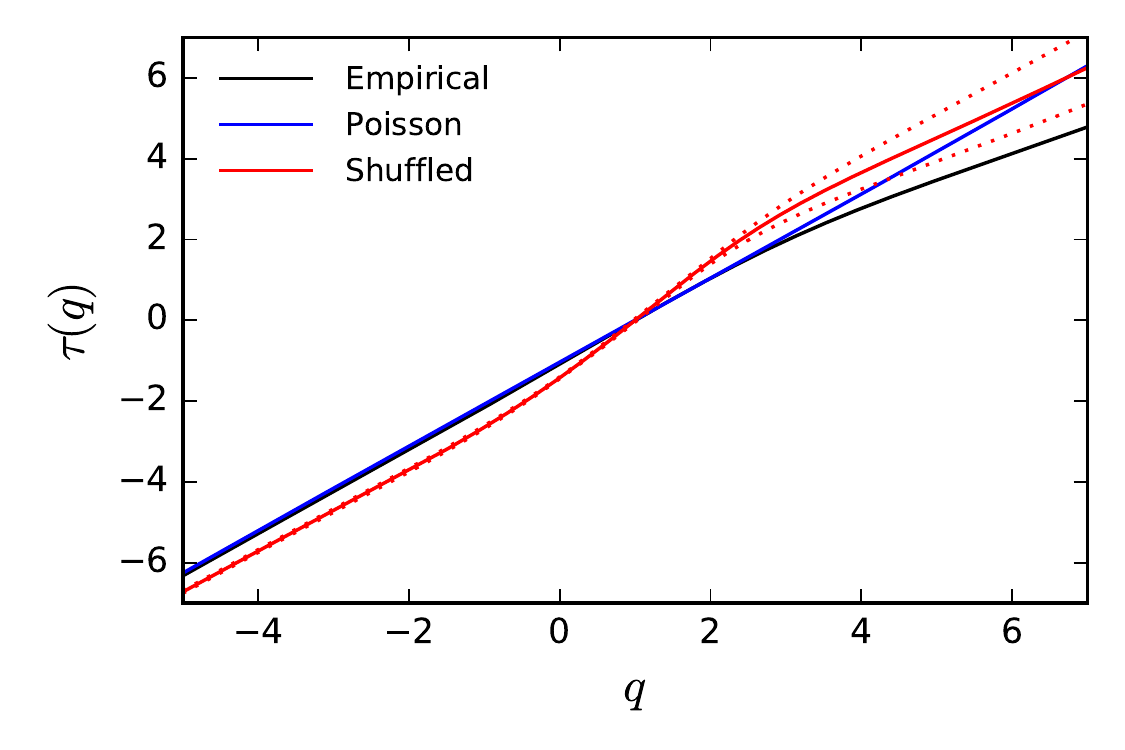}
\caption{Comparison of three particular types of R\'enyi scaling exponent $\tau (q)$ vs. $q$. (i) The black solid curve (taken from Fig.~\ref{figure:Hurstqj}(d)) obtained directly from empirical time series of inter-event times by using our NMF-DFA. (ii) The blue (almost) linearly increasing solid curve was derived from the Poisson distribution. (iii) The solid red curve shows the shuffled empirical inter-event time series. The red dotted curves define its one-sigma error bound, where sigma is a standard deviation.}
\label{figure:tauqcomp}
\end{center}
\end{figure}

As is known, the search for non-spurious/true multifractality first requires the resolution of the role of at least the main factors: (i) main non-stationarity, (ii) finite size effect, and (iii) the broadened distribution. The detrending procedure described in Sec.~\ref{section:ifit} solved point (i), while points (ii) and (iii) are resolved in Fig.~\ref{figure:tauqcomp}, where the R\'enyi scaling exponent $\tau (q)$ is presented vs. $q$ for three characteristic cases.

The blue, almost linearly increasing solid curve (shown in Fig.~\ref{figure:tauqcomp}) was obtained from the Poisson distribution. For this distribution, we have drawn several transactions in each time interval (numbered by index $i$ for each day $\nu $). Based on this, the local mean time of inter-event times between them, $\overline{\Delta t_i^{\nu }}$, was determined (see Fig.~\ref{figure:scheme} for a detailed analysis). These local mean times formed a multi-day time series of length equal to the whole empirical time series of inter-event times. We achieved this by introducing a limitation that the last element (inter-event time) of the time series must be truncated so that the entire synthetic time series is equal to the number of days $N_d$ multiplied by the length of a single session $s\cdot \Delta $.

The presence of possible spurious multifractality here is caused only by the finite size of the time series of inter-event times of the same size as the empirical time series. The spurious multifractality of the Poisson time-series caused only by finite-size effect is negligible in this case as $\tau (q)$ is almost a linear function of $q$. Therefore, we can also expect the influence of the finite size effect on the real multifractality to be negligible. The finite-size effect for the red curve in Fig.~\ref{figure:tauqcomp} is further addressed below. 

The origin of the solid red curve (in Fig.~\ref{figure:tauqcomp}) needs an explanation. We create it in the following three steps:
\begin{itemize}
\item[(i)] We construct the statistics from the empirical series of inter-event times.
\item[(ii)] A new time series is drawn from the statistics thus built by means of shuffling.
In this way, all-time correlations are destroyed if shuffling is performed sufficiently many times.
\item[(iii)] Finally, the exponent $\tau (q)$ is determined from this shuffled time series by our NMF-DFA method.
\end{itemize}

The black solid curve (presented in Fig.~\ref{figure:tauqcomp}) obtained from the empirical time series of inter-event times by using the NMF-DFA, is sufficiently nonlinear to generate multifractality. However, in Fig. 8, we are unable to see the subtle effects in the empirical data. This can only be seen in Fig. 5(a),(e), and (f) in the form of non-monotonic curves, since they depict the derivative $\frac{d\tau (q)}{dq}$.
 
Thus we suggest, by using the NMF-DFA, that an empirical series of the inter-event times yields a true multifractal located far beyond the finite size component and other multifractal pollutions. We suggest that the long-term autocorrelations between absolute values of detrended inter-event time profiles caused the real multifractality in this. These autocorrelations create some true antipersistent structure of fluctuations' clusters of the inter-event times. They are clearly seen in Fig.~\ref{figure:profile}(c) and Fig.~\ref{figure:Stationar}(a), defining the volatility clustering effect. Interestingly, intraday empirical data are sufficient to detect true multifractality, even though the autocorrelations of the inter-event times mentioned are long-term, stretching for many days.

This work is based on two main pillars. First of all, on the NMF-DFA approach constructed in work, which was inspired by the canonical MF-DFA. Using the NMF-DFA based approach, we have here proved that the time series of inter-event times can have a multi-branched multifractal character. Secondly, we have demonstrated that this type of multifractality can lead to phase transitions of the first and second orders according to the Mandelbrot classification. We want to draw attention to the high similarity of both phase transitions to the corresponding phase transitions of the first and second orders according to the Ehrenfest classification.

In the case of traditional multifractality, the phase transition of the first order disappears, which reduces the area of metastable and unstable phases to zero. This implies that canonical multifractality corresponds to critical or supercritical states of the system. 
Multifractality presented in this work is subcritical, where stable, metastable, and unstable phases are all present. From the perspective of this work, traditional multifractality can be regarded as only one of several classes of the full classification. Therefore, the concept of multifractality has been substantially broadened.

\begin{acknowledgments}
The authors are grateful for stimulating discussions with S. Dro\.zd\.z, D. Grech, and P. O\'swi\c{e}cimka.
\end{acknowledgments}

\appendix

\section{Remarks on pre-processing and coarse-graining}\label{section:AppendA}

We compare the pre-processing of our approach with the seminal work by Kantelhardt et al.~\cite{KZKBHS}, emphasizing the differences between the two. 

In both approaches the number of points or empirical data is determined once. The length of the time series $N$ is measured herein by the product $T\cdot N_d=N$, where segment $T$ is the fixed daily length of the time series (or duration of the session measured by time) and $N_d$ is the set number of days (or sessions). The scale $s$, the same for each day, is introduced herein by equality $T=s\cdot \Delta $, where $\Delta $ is the length of the time window or sub-segment and $s$ the number of these sub-segments within the segment. Thus, two separated timescales are present here: one, a more microscopic defined by $s$ and the other one by $N_d$ (cf. Fig.~\ref{figure:scheme}). 

Above we give a different approach than that presented in Kantelhardt et al. work~\cite{KZKBHS}, where a single timescale $s$ is present for a given number of segments or windows $n$. That is, $s=N/n$, where $s$ and $N$ define herein the length of the window or segment and the entire time series, respectively, measured in the number of empirical data points and not in time. Apparently, $s$ introduced herein is different from the one used by us.

Our approach distinguishes between individual sessions (each with the same duration of $T$) as opposed to the method of Kantelhardt et al.~\cite{KZKBHS}. This distinction is natural, and we cannot ignore it. For example, our approach allows to distinguish possible jump of quotation at the opening of each session. In general, this may be different from intra-sessional (intraday) jumps. The approach of Kantelhardt et al. equates inter-sessional with intra-sessional jumps -- it does not make it possible to distinguish them. It loses a piece of information that is perhaps important, and we cannot mask it. It has its consequences in multi-scale divisions of the time series.

Fig. 1 in Kantelhardt et al. and Fig. 1 in our work illustrate the way of introducing scales in both approaches. In the approach of Kantelhardt et al. there is a division of the time series into segments of length $s$ each. Hence, there is $N_s=[N/s]$ such segments, where $[\ldots ]$ means truncation to a natural number. These segments are detrended individually, which leads (most often) to the non-physical jumps of the trend on the borders of subsegments. 

Our approach is devoid of the above-mentioned disadvantage, as detrending is conducted for each session separately. The possible trend jump at the opening of each session here is consistent with the spirit of quotations allowing such increases. No other trend jumps occur in our approach. There is no separate detrending inside any time window $\Delta $, only detrending the entire session. Such an approach does not destroy or remove any fluctuations, nor does it produce artifacts in the form of the trend changes.

The way the scale is introduced is the crucial technical element of both approaches. As one can see, we enter the concrete scale through time window $\Delta $ or (equivalently) through the number $s$ of time windows. Inside each window $\Delta $, we build the time average of inter-event time intervals. That is, we replace the intra-segment detrending present in the approach of Kantelhardt et al. by an intra-window average value and one-session detrending or intra-day detrending. It now detrends a series of these (local) average values, removing intra-day non-stationarities, e.g., the lunch effect. 

The average of the power-law dependence within the $\Delta $ interval does not change its exponent as long as it is at most a slowly-changing function of variable $x$ and $\Delta \ll x$. These results from the following integration,
\begin{eqnarray}
& &\frac{1}{\Delta }\int_x^{x+\Delta }\frac{1}{y^{1+\alpha }}dy\propto \frac{1}{\Delta }\frac{1}{x^{\alpha }}\left(1-\left(1+\frac{\Delta }{x}\right)^{-\alpha }\right) \nonumber \\
&\approx &\frac{1}{\Delta }\frac{1}{x^{\alpha }}\left(1-\exp\left(-\alpha \frac{\Delta }{x}\right)\right)\approx \frac{\alpha }{x^{1+\alpha }}.
\nonumber \\
\label{rown:power-law} 
\end{eqnarray}

The time series consisting of the average values (built independently in each time window $\Delta $) does not lose fluctuations but only decreases their amplitude by the standard factor $\sqrt{n_i^{\nu}}$.  

\section{Autocorrelation functions}\label{section:appendixB}

\begin{figure*}
\begin{center}
\includegraphics[scale=0.80,angle=0,clip]{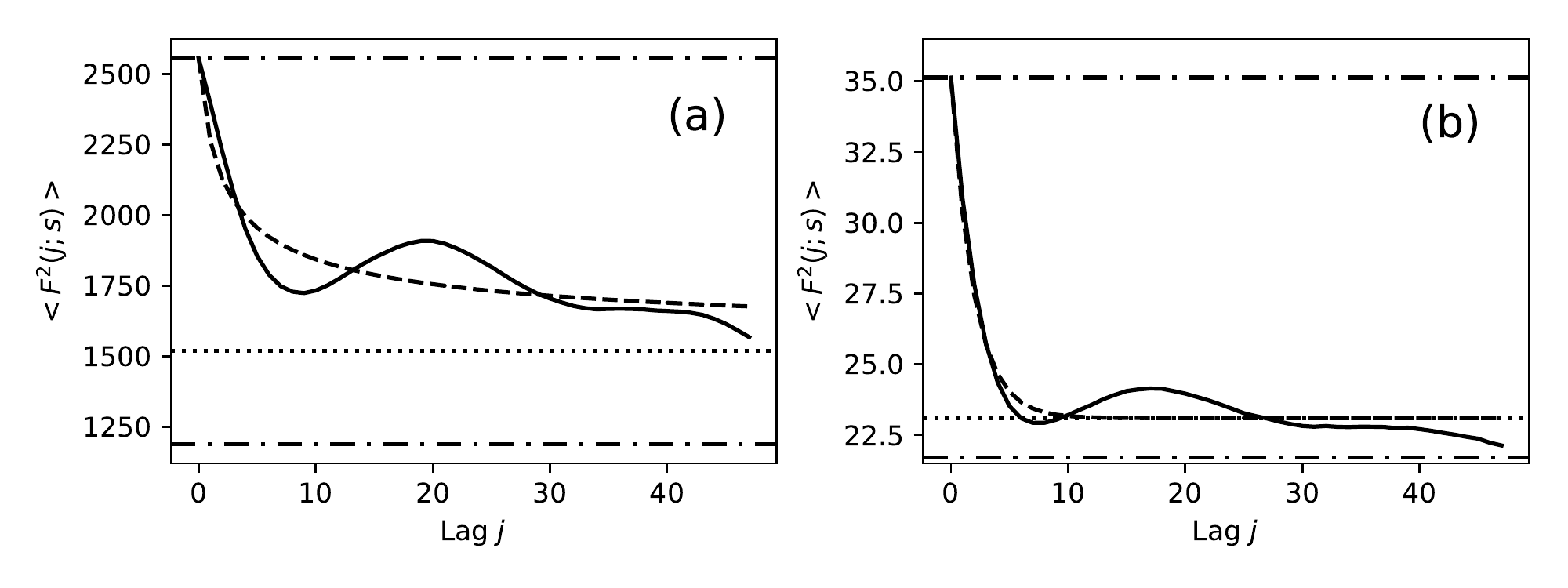}
\caption{The comparison of two intra-day nonlinear autocorrelation functions $\langle F^2(j;s)\rangle $ vs lag $j$ averaged over statistical ensemble of $N_d$ days (for instance, for $s=94$). (a) The power-law (dashed curve) satisfactorily suites the decay of the averaged empirical autocorrelation function $\langle F^2(j;s)\rangle $ (solid curve). Note that the upper dashed-dotted horizontal line represents $\langle (U-y)^2\rangle $, while the bottom one $\langle U-y\rangle ^2$. The dotted horizontal line represents the vertical, constant shift of the power-law. Presumably, location of this autocorrelation function above $\langle U-y\rangle ^2$ is due to the existence of a pattern within the time series of inter-event times (see Fig.~\ref{figure:profile} for details). The plot (b) shows the decaying of the nonlinear autocorrelation function for the time-series of inter-event times generated by the Poisson process. This process is based on the empirical mean inter-event time separately for each day. The exponential function is fitted quite well here to the data (dashed curve).}
\label{figure:Stationar}
\end{center}
\end{figure*}

The plots in Fig.~\ref{figure:Stationar} present the results which are meaningful for our further considerations. 

It seems that the empirical autocorrelation function $\langle F^2(j;s)\rangle \stackrel{\rm def.}{=}\frac{1}{N_d}\sum_{\nu =1}^{N_d}F^2(j;\nu ,s)$ shown in plot Fig.~\ref{figure:Stationar}(a) (by solid curve) is a very slowly converging (and waving) function. It is roughly approximated by the (shifted) power-law, $\langle F^2(j;s)\rangle = A/(a+j)^{\alpha }+\mbox{const}$ (dashed curve), where the fitted shape exponent $\alpha =0.49\pm 0.43$ is definitely smaller than $1$, fitted amplitude $A=1049\pm 387$, while the background parameter $\mbox{const} = 1519\pm 234 > \langle U-y\rangle ^2 = 1190$. Hence, shift parameter $a=A^{1/\alpha }\left(\langle (U-y)^2\rangle -\mbox{const}\right)^{-1/\alpha }=1.0\pm 0.90$, where $\langle (U-y)^2\rangle =2556$. This expression for the shift parameter is vaild because we used equality $\langle F^2(j=0;s)\rangle =\langle (U-y)^2\rangle $, which we have directly from Eq. (\ref{rown:F2snj}) and definition of $\langle \ldots \rangle $.

The slow convergence of the autocorrelation function to positive values result from its construction based on absolute values of deviations (fluctuations), which are always non-negative. In addition, its wavy behavior contains some information about the existence of a long-term fluctuation structure. We have grounds to suppose that this structure is the result of the presence of the long-range correlations between fluctuations -- they are the reason for the creation of this structure and not the other way round.

The autocorrelation function for the canonical Poisson process is presented (solid curve) in Fig.~\ref{figure:Stationar}(b) as a reference case. The exponential function (dashed curve) $g(j)=A\exp(-a\cdot j)+\mbox{const}$ well fits the data, where $A=12.06\pm 0.08,~a=0.496\pm 0.031,~\mbox{const}=23.12\pm 0.08$. The fact that the dotted line does not coincide with the dashed-dotted line is considered to be a manifestation of the finite size effect. Because the relative difference is of the order of one percent, we have reason to believe that in the case of Fig.~\ref{figure:Stationar}(a) (where this difference is of the order of ten percent) the role of this effect is negligible. The solid curves presented in plots in Fig.~\ref{figure:Stationar} represent average values (over the statistical ensemble of days). Therefore, these curves obtained with higher accuracy.

\section{Partial partition functions}\label{section:appendixC}

We start this section with the introduction of reduced (relative) auxiliary quantities. It helps us define the partial partition functions that are crucial to this work.

By substituting both equalities in Eq. (\ref{rown:pjnus}) to Eq. (\ref{rown:ZqsjD}) and using scaling hypothesis given by Eq. (\ref{rown:scalZ}) supported by Eq. (\ref{rown:scaNN}) we get, 
\begin{eqnarray}
Z_q(s)&\approx &\frac{1}{N_d^{q-1}}A^{rel}_qs^{qh^{rel}(q)}=\frac{1}{N_d^{q-1}}A^{rel}_qs^{\tau^{rel}(q)} \nonumber \\
&=&\frac{1}{N_d^{q-1}}A^{rel}_qs^{(q-1)D^{rel}(q)},
\label{rown:scaNZ}
\end{eqnarray}
where the relative (or reduced) prefactor $A_q^{rel}\stackrel{\rm def.}{=}A_q/(A_{q=1})^q$ hence $A_{q=0}^{rel}\approx 1$, $A_{q=1}^{rel}=1$, the relative (or reduced) generalized Hurst exponent $h^{rel}(q)\stackrel{\rm def.}{=}h(q)-h(q=1)$ is vanishing at $q=1$, and the relative (or reduced) scaling exponent $\tau^{rel}(q)\stackrel{\rm def.}{=}qh^{rel}(q)$ is vanishing at $q=0$ and $1$. Having a well-defined reduced scaling exponent, we introduce a formal analog of R\'enyi dimensions, $D^{rel}(q)=\tau^{rel}(q)/(q-1)$, that is relative (reduced) ones, vanishing at $q=0$. As you can see, in this representation you do not need any information about the Hausdorff dimension of the time series support. The information $D^{rel}(q=1)$ and correlation $D^{rel}(q=2)$ dimensions have formally the same forms as the corresponding canonical R\'enyi dimensions (see Appendix~\ref{section:appendixF} for details).

Finally, we can write Eq. (\ref{rown:scaNZ}) in the form of the product of the partial partition functions, key for our further considerations,
\begin{eqnarray}
Z_q(s)\approx \frac{1}{N_d^{q-1}}A_q^{rel} s^{\tau ^{rel}(q)}=Z_q^{lin}(s)\tilde{Z}_q(s),
\label{rown:scaNZfin}
\end{eqnarray}
here
\begin{eqnarray}
Z_q^{lin}(s)&=&\frac{1}{N_d^{q-1}}A^{rel}_qs^{-(q-1)D(q=0)}, \nonumber \\
\tilde{Z}_q(s)&=&s^{\tau (q)},
\label{rown:ZZq}
\end{eqnarray}
where we have to define 
\begin{eqnarray}
D(q=0)\stackrel{\rm def.}{=}h(q=1) 
\label{rown:Dq0}
\end{eqnarray}
for self-consistency (without more profound analysis of this fact herein), while scaling exponent 
\begin{eqnarray}
\tau (q)\stackrel{\rm def.}{=}qh(q)-D(q=0). 
\label{rown:tauq}
\end{eqnarray}
The quantity $D(q=0)$ requires a comment.

The factorization of the partition function given by Eq. (\ref{rown:scaNZfin}) makes it possible to extract the partial partition function $\tilde{Z}_q$ from it. This partial partition function is the most significant because it contains generalized Hurst exponent $h(q)$, which we consider in Sec.~\ref{section:L-F_transform}. Indeed, $h(q)$ is the multifractality core.

\section{$\alpha (q)$ vs. $h(q)$}\label{section:appendixD}

\begin{figure}
\begin{center}
\includegraphics[scale=0.7,angle=0,clip]{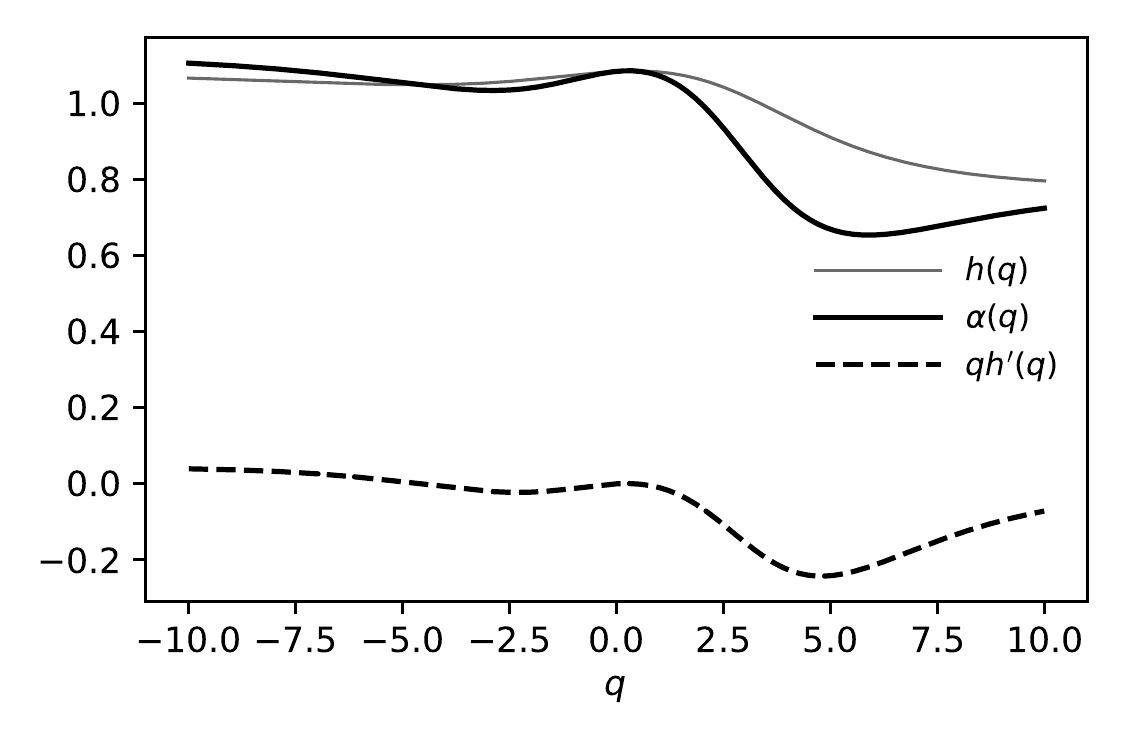}
\caption{Behaviors of $h(q), q\frac{h(q)}{dq}$, and $\alpha (q)$ depending on $q$. The shape similarity of all three curves is clearly visible.}
\label{figure:alphqh}
\end{center}
\end{figure}

In development on Eq. (\ref{rown:alph}) we can answer in detail the question important for further considerations. Namely, does the component $q\frac{dh(q)}{dq}$ appearing in this equation destroy the non-monotonicity imposed by the dependence $h(q)$ on $q$? To this end, we compared on Fig.~\ref{figure:alphqh} the behaviors of $h(q), q\frac{h(q)}{dq}$, and $\alpha (q)$ depending on $q$. As one can see, the non-monotonic dependence of $\alpha $ on $q$ is even enhanced by $q\frac{h(q)}{dq}$ component especially for $q>5$. It is the non-monotonicity of the function $\alpha (q)$ that is directly responsible for the multi-branching of spectrum $f(\alpha )$. The relationship between the non-monotonicity and multi-branchist is discussed in detail in Sec.~\ref{section:fsphasetr}. Here it is enough to note that non-monotonicity and multi-branchist are two sides of the same coin, since the rotation of Fig.~\ref{figure:alphqh} around a bisector of a right angle $180$ degrees gives a bifurcating $\frac{df(a)}{da}(= q(\alpha ))$ curve depending on $\alpha $. Hence, we have multi-branching depicted in Fig.~\ref{figure:cqvsq}(a) and Fig.~\ref{figure:cqvsq}(b), where turning points $C, A_1$, and $B_1$ are the bifurcation points.

We can now precisely define a multi-branch function having continuous derivative bifurcating.

\section{Interpretation of $D(q=0)$}\label{section:appendixE}

In the canonical MF-DFA approach, one can read directly from the scaling relation for the partition function that $D(q = 0)$ is the support's Hausdorff dimension for the time series -- usually $D(q=0)=1$. 

Since in our approach, the R\'enyi dimensions enter into the generalized partition function on a relative way, such a diagnosis does not take place. Therefore, $D(q=0)$ does not have to be a fractal dimension of the substrate (and $D^{rel}(q=0)$ even vanishes). The knowledge of $D(q=0)$ does not require its value to take independently outside our formalism. It is designated by the information Hurst exponent $h(q=1)$ -- it is related to information and not topology. For this reason, the $D(q)$ family should rather be called pseudo R\'enyi dimensions, while the $D^{rel}(q)$ family of the relative or reduced one despite the fact that for $q\neq 0 $ both families have formally the usual interpretation (see Appendix~\ref{section:appendixF} for details). However, in the further part of the work, we return to the simplified name `R\'enyi dimensions' for $D(q)$ of course, remembering the above-given conditions.

Using the scaling exponent $\tau $, we can define now the R\'enyi dimensions in the usual way
\begin{eqnarray}
D(q)\stackrel{\rm def.}{=}\frac{\tau (q)}{(q-1)},
\label{rown:tauZqsjD}
\end{eqnarray}
which additionally allows to present $D^{rel}$ in the reduced form $D^{rel}(q)=D(q)-D(q=0)$. Notably, all the relative quantities defined above (and indexed by $\mbox{`rel'}$) disappear in either $q=0$ or/and in $q=1$, which results from their relative character. 

The partial partition functions $Z_q^{lin}$ and $\tilde{Z}_q$ are normalized separately, and the factorization given by Eq. (\ref{rown:scaNZfin}) (up to multiplicative prefactor and additive exponents) is unique. These partition functions represent statistically independent monofractal and multifractal structures, respectively. We pay attention to the most interesting, the latter one.

\section{Properties of the multi-branched multifractal}\label{section:appendixF}

In this section, we consider the chosen characteristics of multifractality at some significant values of~$q$.

\subsection{Case $q \rightarrow 0$}

This case fundamentally distinguishes our multi-branched multifractality from the ordinary single-branched multifractality. Our approach is unified -- it is entirely based on the generalized Hurst exponent.
From Eqs. (\ref{rown:ZqsjD}), (\ref{rown:scaNZfin}) -- (\ref{rown:tauq}) we obtain $\tau(q=0)=-D(q=0)=-h(q=1)$ (see also Fig.~\ref{figure:Hurstqj}(d) for details), where we took advantage of the fact that generalized Hurst exponent is finite. You can see that the scaling exponent is controlled at $q=0$ only by the generalized Hurst exponent at $q=1$, which has nothing to do with the support of the time series.

\subsection{Case $q \rightarrow 1$: Shanon information} 

In this case one can write the expansion,
\begin{eqnarray}
\tau (q)&\approx&(q-1)[h(q=1)+q\frac{dh(q)}{dq}\mid _{q=1}\nonumber \\
&+&\frac{1}{2}q(q-1)\frac{d^2h(q)}{dq^2}\mid _{q=1}], 
\label{rown:tauqhD1}
\end{eqnarray}
based on the expansion of $h(q)$ in the vicinity of $q=1$, where the expression in square brackets is indeed, 
\begin{eqnarray}
D(q)&\approx &D(q=0)+q\frac{dh(q)}{dq}\mid _{q=1} \nonumber \\
&+&\frac{1}{2}q(q-1)\frac{d^2h(q)}{dq^2}\mid _{q=1}
\label{rown:Dq}
\end{eqnarray}
that is, the expansion of R\'enyi dimensions in the vicinity of $q=1$.

Equivalently we have,
\begin{eqnarray}
\tau ^{rel}(q)&\approx&(q-1)[q\frac{dh(q)}{dq}\mid _{q=1}\nonumber \\
&+&\frac{1}{2}q(q-1)\frac{d^2h(q)}{dq^2}\mid _{q=1}], 
\label{rown:tauqhD1rel}
\end{eqnarray}
where the expression in square brackets is in fact,
\begin{eqnarray}
D^{rel}(q)&\approx &q\left[\frac{dh(q)}{dq}\mid _{q=1} 
+\frac{1}{2}(q-1)\frac{d^2h(q)}{dq^2}\mid _{q=1}\right]. \nonumber \\
\label{rown:Drelq}
\end{eqnarray}
Expansion (in the vicinity of $q=1$) in Eq. (\ref{rown:Drelq}) emphasizes that $D^{rel}(q)$ depends on the successive derivatives of the generalized Hurst exponent as parameters (calculated at $q=1$).

For instance, combining Eqs. (\ref{rown:ZqsjD}) with (\ref{rown:scaNZfin}), we obtain an expression,
\begin{eqnarray}
D^{rel}(q=1)&=&\frac{1}{\ln s}\sum_{\nu =1}^{N_d}p(\nu ,s)\ln p(\nu ,s) \nonumber \\
&=&\frac{1}{\ln s}\langle \ln p(\nu ,s)\rangle = \frac{1}{\ln s}I_{q=1}(s), \nonumber \\ 
\mbox{or equivalently} \nonumber \\
D(q=1)&=&D(q=0)+\frac{1}{\ln s}I_{q=1}(s), 
\label{rown:Dq1}
\end{eqnarray}
here $\langle \ldots \rangle = \sum_{\nu =1}^{N_d}p(\nu ,s)\ldots $ and $I_{q=1}(s)$ can be identified with the Shanon information (within the scale of $s$).

Finally, from Eqs. (\ref{rown:Drelq}) and (\ref{rown:Dq1}) we get,
\begin{eqnarray}
\frac{1}{\ln s}I_{q=1}(s)=\frac{dh(q)}{dq}\mid _{q=1}
\label{rown:AppIq1}
\end{eqnarray}
for $s$ from the scaling region. Thus, the change of the generalized Hurst exponent at $q=1$ is the key to the Shanon information.

\subsection{Case $q\rightarrow 2$}

The correlation integral (or autocorrelation function) is obtained from the $q$-correlation function by substituting $q=2$. Grassberger and Proccacia introduced both quantities long ago~\cite{GP}. They proved that the statistical sum given by Eq. (\ref{rown:ZqsjD}), transforms into a $q$-correlation function. From Eq. (\ref{rown:scaNZ}) we get (for large $s$ for the scaling region) the reduced correlative dimension in the form,
\begin{eqnarray}
D^{rel}(q=2)&\approx &\frac{\ln Z_{q=2}(s)}{\ln s} \nonumber \\
\mbox{or equivalently} \nonumber \\
D(q=2)&\approx &D(q=0)+\frac{\ln Z_{q=2}(s)}{\ln s}.
\label{rown:AppDrel}
\end{eqnarray}

\subsection{General case of arbitrary $q$: Bounds} 

\subsubsection{Properties of $D$}

In our situation (see Eq. (\ref{rown:tauZqsjD}) and Fig.~\ref{figure:Hurstqj} for help) R\'enyi dimensions fulfill general inequalities/bounds which are not identical to those well known for the ordinary R\'enyi dimensions. The differences result from the fact that $D(q)$ is not in our case the monotonic function of $q$ (see Fig.~\ref{figure:Hurstqj}(e) for details), i.e., the Hentschel-Procaccia inequality~\cite {HP} is valid in our case only on disjoint intervals $q$. These bounds are as follows,
\begin{itemize}
\item[(i)] $D(q)>0$, for arbitrary value of $q$;
\item[(ii)] $(q^{\prime }-1)D(q^{\prime })>(q-1)D(q)$ for $q^{\prime }>q$;
\item[(iii)] if $D(q^{\prime })<D(q)$ for $q^{\prime }>q$ (i.e., if we deal with monotonically decreasing ranges of $D(q)$) then $\frac{q^{\prime }-1}{q^{\prime }}D(q^{\prime })>\frac{q-1}{q}D(q)$, where $q^{\prime },q\neq 0$, otherwise the opposite inequality is fulfilled.
\end{itemize}
From (iii) we obtain,
\begin{itemize}
\item[(a)] $D(q)<\frac{q}{q-1}D(q=+\infty )$~for $q>1$;
\item[(b)] $D(q)>\frac{q}{q-1}D(q=-\infty )$~for $q<0$, where $D(q=-\infty )$ is finite.
\end{itemize}

\subsubsection{Properties of $f$}

We begin with general useful property of the $f(\alpha (q))$ spectrum. From Eqs. (\ref{rown:tauZqsjD}) and (\ref{rown:fqalpha})
\begin{eqnarray}
f(\alpha (q))=D(q)+q(q-1)D^{\prime }(q),
\label{rown:Appfaq}
\end{eqnarray}
where we marked $D^{\prime }(q)=\frac{dD(q)}{dq}$.
Note that the intermediate step in the derivation of the above formula is the following convenient expression obtained from Eq. (\ref{rown:tauZqsjD}) and the first equality in Eq. (\ref{rown:fqalpha}),
\begin{eqnarray}
\alpha (q)=D(q)+(q-1)D^{\prime }(q).
\label{rown:Appaq}
\end{eqnarray}

Hence, for $q=1$ and extrema of $D(q)$ we have,
\begin{eqnarray}
\alpha (q)=D(q).
\label{rown:Appa1extr}
\end{eqnarray}
However, you have to see that the location of the extremes of the functions $\alpha (q)$ and $D(q)$ is different (see Figs.~\ref{figure:Hurstqj}(e) and~\ref{figure:Hurstqj}(f) for details).

Moreover, we present two characteristic limitations that can significantly distinguish multi-branched multifractality from ordinary (i.e., single-branched) multifractality. Namely, from Eq. (\ref{rown:Appa1extr}) we obtain
\begin{eqnarray}
\alpha ({q=+\infty })=D(q=+\infty ) \nonumber \\
\alpha ({q=-\infty })=D(q=-\infty ),
\label{rown:Applim}
\end{eqnarray}
on assumption that derivative $D^{\prime }(q)$ disappears faster than $1/q$  if $\mid q\mid \rightarrow \infty $. It should be emphasized that because $\alpha (q)$ is not a monotonically decreasing function of $q$ (see Fig.~\ref{figure:Hurstqj}(f) for details), in general $\alpha (q=+\infty )\neq \alpha _{min}$ and $\alpha (q=-\infty )\neq \alpha _{max}$, where 
$\alpha _{min}~\mbox{and}~\alpha _{max}$ are the minimal and maximal values of $\alpha (q)$, respectively. 

From Eq.~\ref{rown:Appfaq} the special cases yield,
\begin{eqnarray}
f(\alpha (q=0))&=&D(q=0)=\alpha (q=0)+D^{\prime }(q)\mid _{q=0}, \nonumber \\
f(\alpha (q=1))&=&D(q=1)=\alpha (q=1),
\label{rown:App01}
\end{eqnarray}
where we obtain $\frac{df(\alpha (q))}{d\alpha }\mid _{\alpha (q=1)}=1$ (with help of Eq. (\ref{rown:dfqalpha})). This relation and the second equality in Eq. (\ref{rown:App01}) defines the contact point considered in Sec.~\ref{section:L-F_transform}. 

\section{Remarks on Legendre-Fenchel transformation}\label{section:appendixG}

We present the essence of the one-dimensional Legendre-Fenchel (LF) transformation in the version we use in our approach. In this way, we want to make our work more accessible/understandable. The LF transformation has never been used in a multifractality context. We show that it can be beneficial by significantly expanding the possibilities of advanced time series analysis. This is why we want to draw it to the attention of physicists, especially those dealing with complex systems, especially in econo- and sociophysics.

\begin{figure}
\begin{center}
\includegraphics[scale=0.7,angle=0,clip]{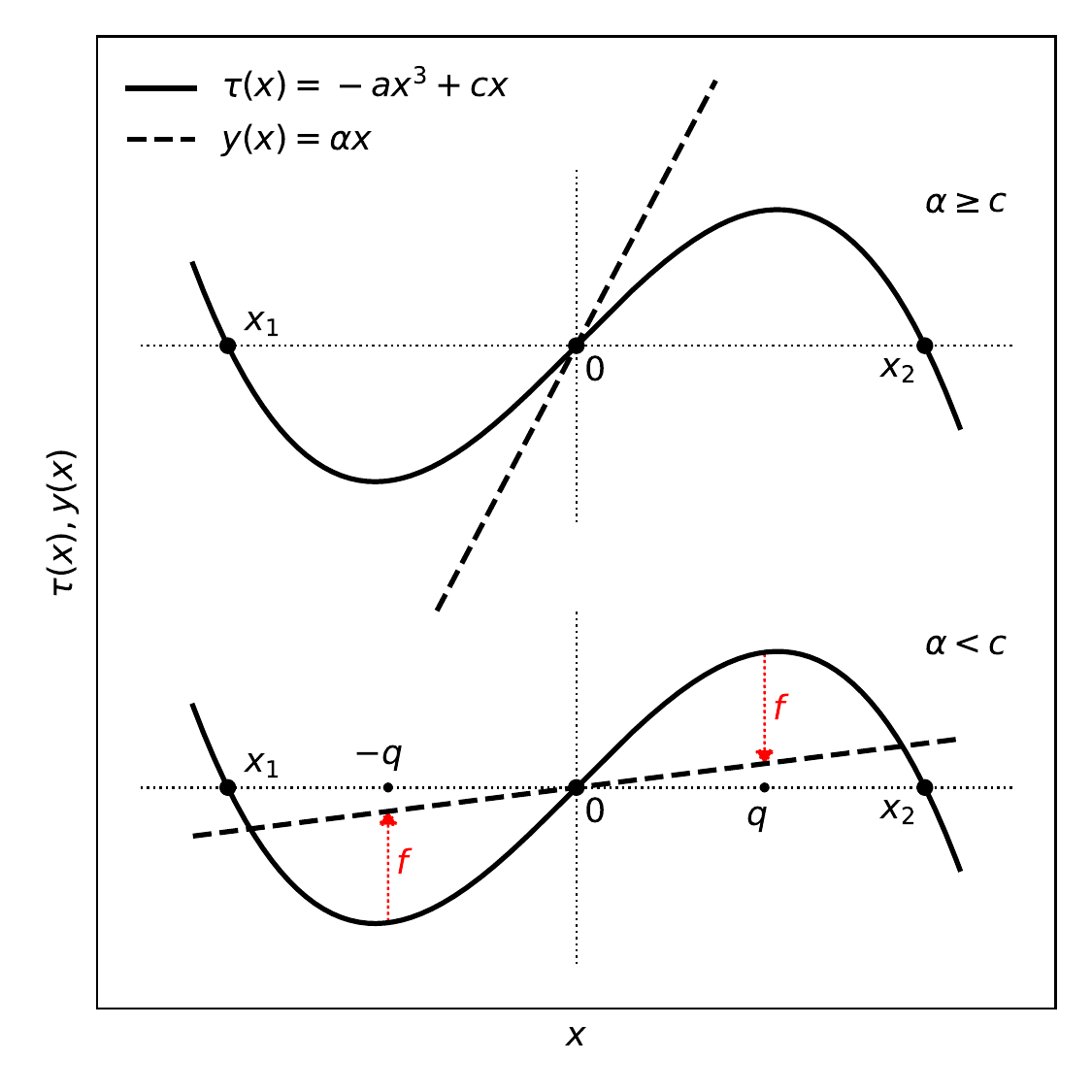}
\caption{The schematic plots of convex-concave function $\tau (x)$ (solid curve on both plots) with two supremes denoted on the bottom plot by $f$ (see Eq. (\ref{rown:fxSupl}) for details). This solid curve shape is the basis for further consideration.}
\label{figure:FigX1}
\end{center}
\end{figure}

The Legendre-Fenchel transformation can refer, for example, to functions whose schematic exemplary course we show in Fig. \ref{figure:FigX1}. As can be seen, it alternately contains both convex and concave parts and not just one of them, as in the case of the Legendre transformation. Thus, the LF transformation allows treating both stable and metastable states, for example, in the physics of phase transitions. This is why we are dealing with two supremes of the function 
\begin{eqnarray}
\mid f(x)\mid =\mid y(x)-\tau (x)\mid 
\label{rown:fxSupl}
\end{eqnarray}
for the fixed $\alpha $ (see the bottom plot in Fig. \ref{figure:FigX1}) located at points $-q$ and $q$. We obtain these supremes as roots of the equation 
\begin{eqnarray}
\frac{df}{dx} = 0 = 3a q^2-(c-\alpha ), 
\label{rown:dfdx}
\end{eqnarray}
where the variable $x$ runs the entire domain while the variable $q$ applies only to supreme. Thus, 
\begin{eqnarray}
q(\alpha )=\mp \sqrt{\frac{c-\alpha }{3a}}, 
\label{rown:mpq}
\end{eqnarray}
where coefficient $\alpha $ is the directional coefficient of the straight line $y(x) = \alpha x$. The coefficient $c$ of the basic polynomial 
\begin{eqnarray}
\tau (x) = - a x^3+c x
\label{rown:taux}
\end{eqnarray}
sets the upper limit of $0\leq \alpha < c$ below, which these supreme exist and are located in the range $x_1\leq -q$ and $q\leq x_2$. The variable $q$ denotes the value of the variable $x$ at the extreme point of the function $f(x)$. Of course, the size of $q$ depends on the slope $\alpha $ of the straight line $y(x)$. If the component containing $x^2$ were present in the $\tau (x)$ polynomial, then the extremes of this polynomial would not be located symmetrically on both sides of the vertical axis. Such a situation would be more complicated, bringing nothing new to the understanding of the essence of the functioning of the LF transformation. 

From Eq. (\ref{rown:mpq}) we obtain directly an inverse expression,
\begin{eqnarray}
\alpha (q)=-3aq^2+c
\label{rown:alphaq}
\end{eqnarray}
and finally (using Eq. (\ref{rown:fqalpha})) the spectrum of dimensions,
\begin{eqnarray}
f(\alpha )=\pm 2a \left(\frac{c-\alpha }{3a}\right)^{3/2}.
\label{rown:faqFL}
\end{eqnarray}

\begin{figure}
\begin{center}
\includegraphics[scale=0.7,angle=0,clip]{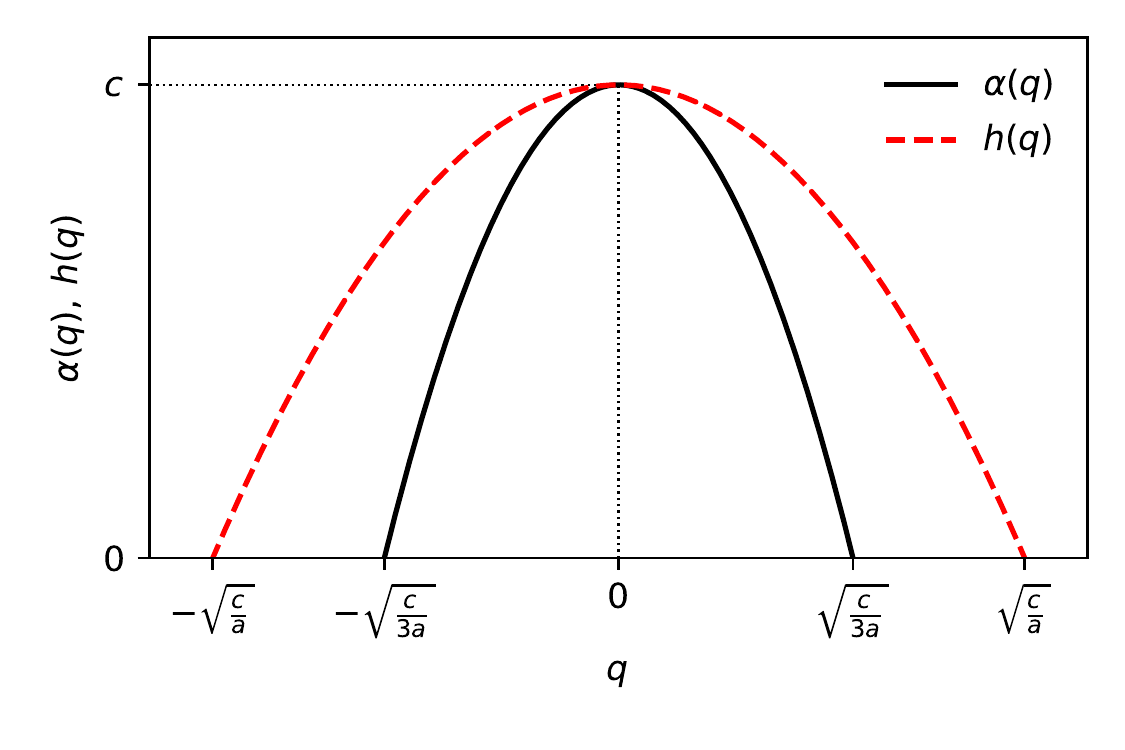}
\caption{Dependence $\alpha $ (solid curve) and $h$ (dashed curve) on $q$ representing Eqs. (\ref{rown:alphaq})  and (\ref{rown:hqSupl}), respectively. The key feature for further considerations is the non-monotonic $\alpha $ and $h$ dependence on $q$.}
\label{figure:FigX2}
\end{center}
\end{figure}

\begin{figure}
\begin{center}
\includegraphics[scale=0.7,angle=0,clip]{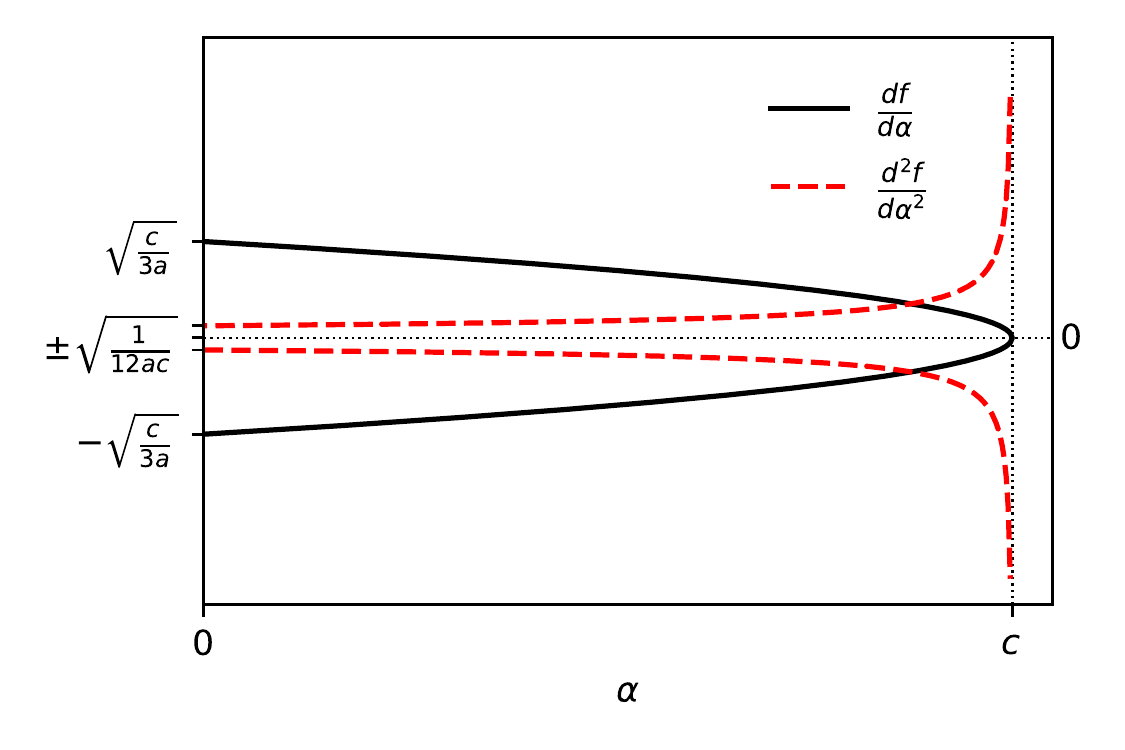}
\caption{Derivatives $\frac{df(\alpha )}{d\alpha }(=q(\alpha )$, solid curve) and $\frac{d^2f(\alpha ) }{d\alpha ^2}$ (dashed curves) vs. $\alpha $. This black bifurcating curve defines the phase diagram and these dashed diverging curves constitute its characteristics, especially at point $\alpha = c$.}
\label{figure:FigX3}
\end{center}
\end{figure}

\begin{figure}
\begin{center}
\includegraphics[scale=0.5,angle=0,clip]{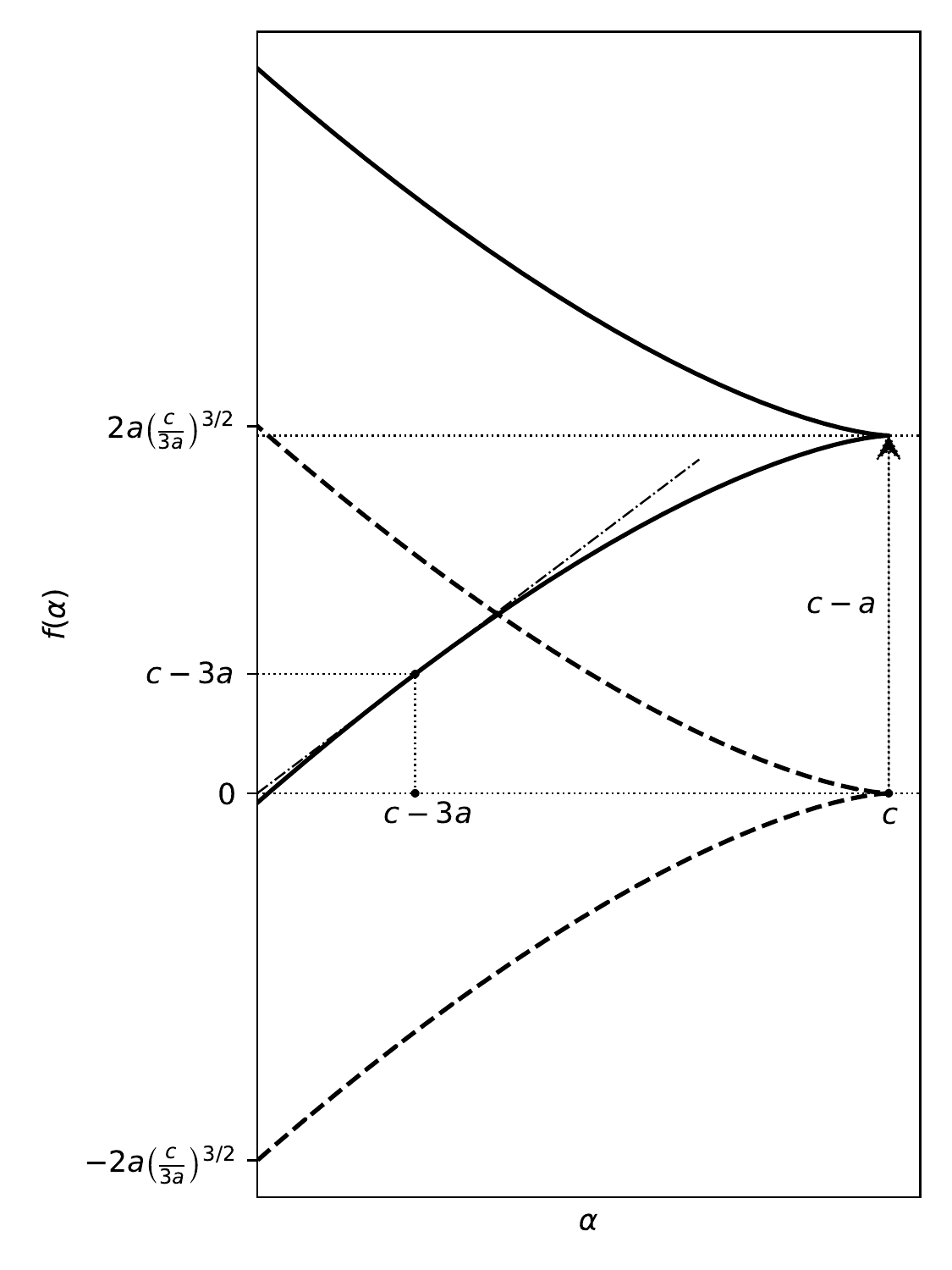} 
\caption{The bifurcating spectra of dimensions $f(\alpha )$ vs. $\alpha $ (solid and dashed two-branched curves). The solid curve has been moved up by $c-a$ relative to the dashed curve. This results in the appearance of the contact point satisfying the equality $f(\alpha)=\alpha =c-3a$, which is the case for $q = 1$. It justifies using the concept of contact transformation in the case of LF transformation. Hence, the maximum value of $f$ or $f(\alpha (q = 0)$ may be greater than 1. However, this does not prevent the additional interpretation of $f$ as the probability density of the variable $\alpha $. It prevents $D (q = 0)$ from being interpreted as a box dimension or capacity.}
\label{figure:FigX4}
\end{center}
\end{figure}

The significant components/functions of the LF transformation given by Eqs. (\ref{rown:mpq}) -- (\ref{rown:faqFL}) are plotted in Figs. \ref{figure:FigX1} -- \ref{figure:FigX4} by the corresponding curves (the plots we made, for example, for $a=1$ and $c=4$). These functions have all the most important features of their empirical counterparts considered in this paper. It is about non-monotonicity, two-branching and left-sided.

Still to meet additional empirical condition $\tau (q = 1)=0$ prompted by Eqs. (\ref{rown:Dq0}) and (\ref{rown:tauq}), we go to the shifted polynomial $\tau (x)$,
\begin{eqnarray}
\tau (x) \Rightarrow \tau (x) + a - c = -ax^3+cx+a-c. \nonumber \\
\label{rown:tauq1Sup}
\end{eqnarray}
According to Eq. (\ref{rown:fxSupl}), this translation shifts the spectrum of dimensions as follows,  
\begin{eqnarray}
f(\alpha ) \Rightarrow f(\alpha ) + c - a .
\label{rown:faqSupl}
\end{eqnarray}
It is this new shifted spectrum of dimensions that is presented by solid curves in Fig. \ref{figure:FigX4} in the form of two branches. As one can see, it is not only two-branched but also left-sided. This is a basic analogue of our empirical spectrum of dimensions shown in plots Fig. 5(a) and Fig. 5(b) in our work. Please note that the shifted functions used do not change the $q$ and $\alpha $ variables as should be the case. 

It should be emphasized that the characteristic example considered here is particularly simple - each of its elements is presented in a closed, easy to determine analytical form.

\subsection{Significant references to our NMF-DFA}

We are now introducing, as defined by (\ref{rown:tauZqsjD}),
\begin{eqnarray}
D(q)=\frac{\tau(q)}{q-1}=-\frac{a(q^3-1)}{q-1}+c
\label{rown:DqSupl}
\end{eqnarray}
where $\tau(q)$ is already this new shifted function. Hence, we get directly
$D(q=0)=c-a$, $D(q=1)=c-3a$ and $D(q=2)=c-7a$. 

Similarly, using definition (\ref{rown:tauq}) we get, 
\begin{eqnarray}
\frac{\tau(q)+D(q=0)}{q}=h(q)=-aq^2+c.
\label{rown:hqSupl}
\end{eqnarray}
This expression is graphically presented in Fig. \ref{figure:FigX2} by the dashed curve.

\subsubsection{Relative quantities}

In the following we will present the list of relative (reduced) quantities used in NMF-DFA. Based on the relative quantities given in Appendix~\ref{section:appendixC}, we get
\begin{eqnarray}
h^{rel}(q)&=&h(q)-h(q=1)=-a(q^2-1) \nonumber \\
\tau ^{rel}(q)&=&qh^{rel}(q)=-aq(q^2-1) \nonumber \\
D^{rel}(q)&=&\frac{\tau ^{rel}(q)}{q-1}=-\alpha q(q+1).
\label{rown:relSupMat}
\end{eqnarray}
We have provided these quantities in an explicit form because they play an important role in the statistical analysis of the time series that we conducted in our work.

\end{document}